\documentclass[%
 reprint,
 amsmath,amssymb,
 aps,
,onecolumn]{revtex4-2}
\usepackage[utf8]{inputenc} 
\usepackage[T1]{fontenc}    
\usepackage{hyperref}       
\usepackage{url}            
\usepackage{booktabs}       
\usepackage{amsfonts}       
\usepackage{nicefrac}       
\usepackage{microtype}      
\usepackage[section]{placeins}
\usepackage{changepage}
\usepackage{amsmath,amsfonts,amsthm,bm}
\usepackage{pgfplots}
\usepackage{float}
\usepackage{tikz}
\usepackage{caption}
\usepackage{subcaption}
\usepackage{setspace}
\usepackage{amssymb}
\usepackage{dcolumn}
\usepackage{siunitx}
\usepgfplotslibrary{groupplots}
\usetikzlibrary{calc,decorations.markings}
\PassOptionsToPackage{dvipsnames,svgnames}{xcolor}   
\usetikzlibrary{arrows,calc}
 \makeatletter%
\usepackage[toc,page]{appendix}
\usetikzlibrary{3d,calc}

\makeatletter
\newcommand{\vast}{\bBigg@{4}}
\newcommand{\Vast}{\bBigg@{5}}
\makeatother

\tikzset{
  on each segment/.style={
    decorate,
    decoration={
      show path construction,
      moveto code={},
      lineto code={
        \path [#1]
        (\tikzinputsegmentfirst) -- (\tikzinputsegmentlast);
      },
      curveto code={
        \path [#1] (\tikzinputsegmentfirst)
        .. controls
        (\tikzinputsegmentsupporta) and (\tikzinputsegmentsupportb)
        ..
        (\tikzinputsegmentlast);
      },
      closepath code={
        \path [#1]
        (\tikzinputsegmentfirst) -- (\tikzinputsegmentlast);
      },
    },
  },
  mid arrow/.style={postaction={decorate,decoration={
        markings,
        mark=at position .5 with {\arrow[#1]{stealth}}
      }}},
  quart arrow/.style={postaction={decorate,decoration={
        markings,
        mark=at position .15 with {\arrow[#1]{stealth}},
        mark=at position .95 with {\arrow[#1]{stealth}}
      }}},
}

\pgfplotsset{compat=1.17} 
\begin{document}
\preprint{AIP/123-QED}

\title[Planar Coil Optimization in a Magnetically Shielded Cylinder]{Planar Coil Optimization in a Magnetically Shielded Cylinder}
\author{M.~Packer\textsuperscript{1}}
\author{P.~J.~Hobson\textsuperscript{1}}%
\author{N.~Holmes\textsuperscript{1,2}}%
\author{J.~Leggett\textsuperscript{1,2}}%
\author{P.~Glover\textsuperscript{1,2}}%
\author{M.~J.~Brookes\textsuperscript{1,2}}%
\author{R.~Bowtell\textsuperscript{1,2}}%
\author{T.~M.~Fromhold\textsuperscript{1}}%
\affiliation{%
 \textsuperscript{1}School of Physics and Astronomy, University of Nottingham, Nottingham, NG7 2RD, UK. \\
 \textsuperscript{2}Sir Peter Mansfield Imaging Centre, School of Physics and Astronomy, University of Nottingham, Nottingham, NG7 2RD, UK.
}
\date{\today}
\begin{abstract}
Hybrid magnetic shields with both active field generating components and high-permeability magnetic shielding are increasingly needed for various technologies and experiments that require precision-controlled magnetic field environments. However, the fields generated by the active components interact with the passive magnetic shield, distorting the desired field profiles. Consequently, optimization of the active components needed to generate user-specified target fields must include coupling to the high-permeability passive components. Here, we consider the optimization of planar active systems, on which an arbitrary static current flows, coupled to a closed high-permeability cylindrical shield. We modify the Green's function for the magnetic vector potential to match boundary conditions on the shield's interior surface, enabling us to construct an inverse optimization problem to design planar coils that generate user-specified magnetic fields inside high-permeability shields. We validate our methodology by designing two bi-planar hybrid active--passive systems, which generate a constant transverse field, $\mathbf{B}=\mathbf{\hat{x}}$, and a linear field gradient, $\mathbf{B}=(-x~\mathbf{\hat{x}}-y~\mathbf{\hat{y}}+2z~\mathbf{\hat{z}})$, respectively. For both systems, the inverse-optimized magnetic field profiles agree well with forward numerical simulations. Our design methodology is accurate and flexible, facilitating the miniaturization of high-performance hybrid magnetic field generating technologies with strict design constraints and spatial limitations.
\end{abstract}
\maketitle
\section{Introduction}
Tailored, high-precision, low magnetic field environments are required for many applications, devices, and experiments. Examples include magnetic field control in quantum sensing of gravity for underground surveying and mapping~\cite{Wueaax0800, doi:10.1038/s41598-018-30608-1, SnaddenGradiometer, AIQuantumSensors}, magnetic field cancellation for atomic magnetometry~\cite{Shah_2013} with applications in medical imaging such as magnetoencephalography~\cite{nature,BOTO2019116099,TIERNEY2019598,pshwin} and neonatal/fetal magnetocardiography~\cite{doi:10.1002/pd.4976,LEW20172470,doi:10.1161/JAHA.119.013436}, and noise suppression in fundamental physics experiments~\cite{LiangClock,RevModPhys.91.015001,doi:10.1063/1.4886146}. Usually, these systems are enclosed by a shield formed from high-permeability material, which excludes stray external magnetic fields that can limit the accuracy and sensitivity of the measurements. In particular, cylindrical shield geometries are often used because the dimensions and spacings of multiple cylindrical shield layers can be optimized to generate a large interior shielded region~\cite{8887195,doi:10.1063/1.5131250}. The magnetic field interior to the shield can then be adjusted by active field generating components to either cancel background fields further or to define a specific field environment. However, the surrounding passive shielding material deforms the magnetic field profiles generated by the active components, making it hard to design wire patterns which accurately generate specified target magnetic field profiles~\cite{doi:10.1002/9780470268483.app2}.

Boundary element methods (BEMs) can be used to optimize magnetic fields generated by surface currents on a triangular mesh~\cite{Pissanetzky_1992,Mpoole,zetter2020magneticfield,mkinen2020magneticfield} to generate arbitrary target magnetic fields. BEMs are extremely powerful and flexible since they can be used to define active systems with complex geometries inside passive shields. They are, however, limited by computational power with results depending on mesh size and on the distance of the active components from the shielding material. Alternatively, analytical methods for optimizing hybrid active--passive systems are advantageous since they provide intuition and understanding of how magnetic fields are distorted by the presence of high-permeability material. Analytical formulations also provide a fast and efficient route to determine the best system for generating a bespoke user-specified magnetic field profile.

Currently, however, analytical models for hybrid active--passive systems are restricted to a limited number of scenarios. Simple discrete coil geometries have been formulated in cylindrical high-permeability magnetic shields, where the magnetic field is decomposed into azimuthal Fourier modes~\cite{Solenoid1, solenoid2, doi:10.1063/1.1719514} and matched at the shield boundary. Planar high-permeability materials have been incorporated into optimization procedures using the method of mirror images~\cite{jackson,KordyukLevitation} to determine the total magnetic field generated by a static current source inside a magnetically shielded room~\cite{cubefem, 8500866, LIU2020166846, niall1}. Similar work in magnetic resonance imaging (MRI) has investigated the interaction of switched magnetic field gradients with high-conductivity materials used for passive shielding~\cite{TurnerBowley1986}. More recently, solutions for the interaction between a static current source and a finite closed cylindrical high-permeability shield have been formulated for the specific case of cylindrical co-axial surface current geometries~\cite{PhysRevApplied.14.054004}. These formulations incorporate the effect of the shielding material by explicitly solving Maxwell's equations by matching the required boundary conditions on the surface interface to find the total magnetic field generated by the system.

Building on this previous work, here we determine the effect of a closed finite length high-permeability cylinder on the magnetic field generated by an arbitrary static current distribution on an interior circular plane, oriented perpendicular to the axis of the cylinder. We calculate the effect of the high-permeability cylinder on the magnetic field produced by current flow on the plane, which enables us to determine optimal current paths to generate user-specified target fields inside the cylinder. We first derive the vector potential generated by an arbitrary planar current source. Then, we calculate a pseudo-current density induced on the cylindrical surface of the high-permeability material in response to the planar current source, and, hence, derive a Green's function for our system. Finally, we implement a Fourier decomposition of the current paths to calculate the total magnetic field in terms of a set of weighted Fourier coefficients. This formulation allows the incorporation of a quadratic optimization procedure to determine globally optimal designs that generate user-specified target magnetic field profiles. Through the use of this optimization procedure, we design two example bi-planar coil systems optimized for operation inside a finite-length closed high-permeability cylindrical magnetic shield. In both cases, we confirm that our
analytical model agrees well with the result of numerical finite element simulations. Our work extends the range of coil geometries which can be efficiently optimized to generate static user-specified target magnetic fields in the presence of high-permeability materials, expanding design flexibility for systems which require precision-controlled magnetic field environments. 

\section{Theory}
High-permeability magnetic shields, usually constructed from materials such as mumetal, are used to attenuate external background magnetic fields in order to shield magnetically sensitive equipment from spurious signals. These high-permeability materials generate an induced magnetization, $\mathbf{M}$, in response to a incident field at their surface, $S$. This ensures continuity of the parallel and tangential components of the magnetic field, $\mathbf{B}$, and magnetic field strength, $\mathbf{H}$, respectively, at the interface between air and a material. Considering this interface while working in the magnetostatic regime with no surface currents, the boundary conditions take the form
\begin{equation}\label{eq.bperp}
    (\mathbf{B_{\mathrm{mat.}}}-\mathbf{B_{\mathrm{air}}})\cdot \mathbf{\hat{n}}=0 \qquad \textnormal{on $S$},
\end{equation}
and
\begin{equation}\label{eq.btan}
    \left(\frac{1}{\mu_r}\mathbf{B_{\mathrm{mat.}}}-\mathbf{B_{\mathrm{air}}}\right)\wedge \mathbf{\hat{n}}=0 \qquad \textnormal{on $S$},
\end{equation}
where $\hat{\mathbf{n}}$ is the unit vector normal to the boundary and $\mu_r$ is the relative permeability of the material. To design magnetic fields effectively in shielded environments, we need to determine the total field generated by an active coil structure and the high-permeability material. The total magnetic field in free space is related to the magnetic field strength and the magnetization by
\begin{equation}\label{eq.mag}
    \mathbf{B}=\mu_0 (\mathbf{H}+\textbf{M}).
\end{equation} 
The induced magnetization can be formulated in terms of a pseudo-current density $\mathbf{\tilde{j}}$, confined to the material's surface
\begin{equation}\label{eq.boundj}
    \mathbf{\nabla} \wedge \mathbf{M}=\widetilde{\textbf{j}},
\end{equation}
with the magnetic field strength related to a current density $\mathbf{j_c}$, on an active structure, using Amp\`ere's law,
\begin{equation}\label{eq.current}
    \nabla \wedge \mathbf{H} = \mathbf{j_c}.
\end{equation}
By using \eqref{eq.mag}-\eqref{eq.current}, and the relation between the vector potential and the magnetic field, $\mathbf{B}=\boldsymbol{\nabla}\wedge\mathbf{A}$, the total vector potential in free space generated by the system can be cast as the Poisson equation,
\begin{equation}
    \nabla^2 \mathbf{A}=-\mu_0(\mathbf{j_c}+\widetilde{\textbf{j}}),
\end{equation}
resulting in the integral solution in terms of an arbitrary current density
\begin{equation} \label{eq.vectorA}
    \mathbf{A\left(r\right)}=\mu_0 \int_{r'} \mathrm{d}^3\mathbf{r'}\ G(\mathbf{r},\mathbf{r'}) \mathbf{j(r')},
\end{equation}
where $G(\mathbf{r},\mathbf{r'})$ is the associated Green's function for the system~\cite{jackson}.

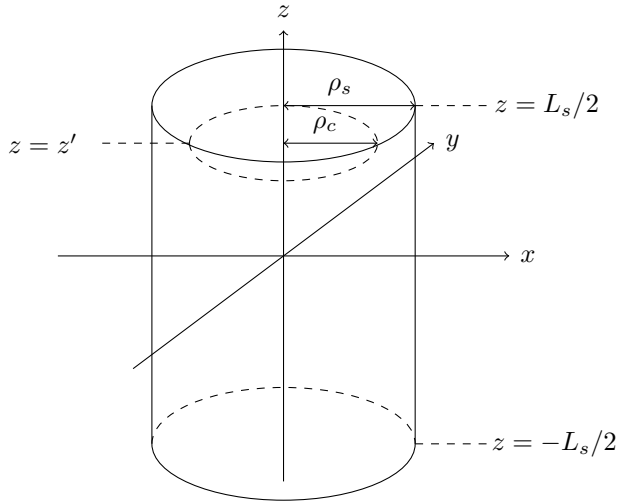
\begin{figure}[htb!]
\begin{center}
\begin{tikzpicture}
\draw [dashed] (-1.25,0) arc (180:360:1.25 and -0.5);
\draw [dashed] (-1.25,0) arc (180:360:1.25 and 0.5);
\draw [->] (-3,-1.5) -- (3,-1.5);
\draw [<-] (0,1.5) -- (0,-4.5);
\draw [->] (-2,-3) -- (2,0);
\draw [<->] (0,0) -- (1.25,0);
\draw [dashed] (-1.25,0) -- (-2.5,0);
\node at (3.25,-1.5) {$x$};
\node at (2.25,0) {$y$};
\node at (0,1.75) {$z$};
\node at (0.55,0.25) {$\rho_c$};
\node at (-3.2,0) {$z=z'$};
\draw (0,0.5) ellipse (1.75 and 0.75);
\draw (-1.75,0.5) -- (-1.75,-4);
\draw (1.75,0.5) -- (1.75,-4);
\draw (-1.75,-4) arc (180:360:1.75 and 0.75);
\draw [dashed] (-1.75,-4) arc (180:360:1.75 and -0.75);
\draw [<->] (0,0.5) -- (1.75,0.5);
\draw [dashed] (1.75,0.5) -- (2.75,0.5);
\draw [dashed] (1.75,-4) -- (2.75,-4);
\node at (0.75,0.75) {$\rho_s$};
\node at (3.5,0.5) {$z=L_s/2$};
\node at (3.6,-4) {$z=-L_s/2$};
\end{tikzpicture}
\end{center}
\caption{A cylindrical hollow high-permeability magnetic shield of length $L_s$, radius $\rho_s$, and with planar end caps located at $z=\pm L_s/2$ encloses an interior circular current source (bounded by the upper dashed curve) of radius $\rho_c<\rho_s$, which lies in the $z=z'$ plane.}
\label{fig.magshield}
\end{figure}

Here, we consider the specific example of a closed cylindrical magnetic shield surrounded by free space, as shown in Fig.~\ref{fig.magshield}. This cylinder has relative permeability, $\mu_r\gg1$, radius $\rho_s$, and length $L_s$, with planar end caps located at $z=\pm L_s/2$. Inside this cylinder, an arbitrary current flows on a circular planar surface of radius $\rho_c<\rho_s$, centered on the $z$-axis and lying in the $z=z'$ plane where $|z'|<L_s/2$. 
Previous work has shown that the magnetic field generated by a high-permeability material in response to an applied static field deviates from that of a perfect magnetic conductor on the scale of $\mathcal{O}\left( \mu_r^{-1}\right)$~\cite{mu, mu1}. More specifically, in the case of a closed finite length cylinder with permeability $\mu_r=20000$, equal to industrial standard mumetal, a material thickness of $d=1$~mm is sufficient to provide a response similar to that of bulk material, differing from the perfect magnetic conducting response only by approximately $0.005\%$  \cite{PhysRevApplied.14.054004}. Therefore, we can approximate sufficiently thick materials with high-permeability $\mu_r>20000$, such as mumetal, to that of a perfect magnetic conductor without compromising the accuracy of a theoretical model. Assuming the surface of the shield shown in Fig.~\ref{fig.magshield} is a perfect magnetic conductor, the boundary condition at its surface can be written as 
\begin{equation}
 B_\rho\bigg\rvert_{z=\pm L_s/2}\hspace{-30pt}=0, \quad\quad B_\phi\bigg\rvert_{z=\pm L_s/2,\rho=\rho_s}\hspace{-50pt}=0, \qquad\quad B_z\bigg\rvert_{\rho=\rho_s}\hspace{-13pt}=0, \label{bbound}
\end{equation} 
where the magnetic field $\mathbf{B}=B_\rho~\boldsymbol{\hat{\rho}}+B_\phi~\boldsymbol{\hat{\phi}}+B_z~\mathbf{\hat{z}}$, is expressed in cylindrical polar coordinates.

As previously formulated for cylindrical current flow~\cite{PhysRevApplied.14.054004}, the response generated by a finite closed cylinder with high-permeability, $\mu_r\gg1$, can be determined by matching the orthogonal modes in the magnetic field and generating a relation between the initial current flow and the resulting magnetization induced on the surface of the cylinder. Here we apply the same methodology to generate an analytical expression for the response of a finite closed high-permeability cylinder to a planar current source using a similar decomposition. The vector potential generated by a current source in cylindrical coordinates can be expressed as,
\begin{equation}\label{eq.arho}
    A_{\rho}\left(\mathbf{r}\right)=\mu_0 \int_{r'} \mathrm{d}^3\mathbf{r'}\ G(\mathbf{r},\mathbf{r'})\left[j_{\rho}\left(\mathbf{r'}\right)\cos\left(\phi-\phi'\right)+j_{\phi}(\mathbf{r'})\sin\left(\phi-\phi'\right)\right],
\end{equation}
\begin{equation} \label{eq.aphi}
    A_{\phi}\left(\mathbf{r}\right)=-\mu_0 \int_{r'} \mathrm{d}^3\mathbf{r'}\ G(\mathbf{r},\mathbf{r'})\left[j_{\rho}\left(\mathbf{r'}\right)\sin\left(\phi-\phi'\right)-j_{\phi}(\mathbf{r'})\cos\left(\phi-\phi'\right)\right],
\end{equation}
\begin{equation} \label{eq.az}
    A_{z}\left(\mathbf{r}\right)=\mu_0 \int_{r'} \mathrm{d}^3\mathbf{r'}\ G(\mathbf{r},\mathbf{r'}) j_{z}(\mathbf{r'}).
\end{equation}
Since there is no current flow in the $z$-direction for a planar current source perpendicular to the axis of the cylindrical shield, the continuity equation can be used to express the planar current flow in terms of a single scalar streamfunction defined on the planar surface
\begin{equation}\label{eq.streamyd}
 j_{\rho}(\mathbf{r'})=\frac{1}{\rho'}\frac{\partial \varphi(\mathbf{r'})}{\partial \phi'}, \qquad\quad j_{\phi}(\mathbf{r'})=-\frac{\partial \varphi(\mathbf{r'})}{\partial \rho'},
\end{equation}
where $\varphi(\mathbf{r'})=\varphi(\rho',\phi')$. To exploit the radial symmetries of the system, we choose to decompose the Green's function in cylindrical coordinates in terms of Bessel functions of the first kind,
\begin{equation}\label{green1}
    G(\mathbf{r,r'}) =\frac{1}{4\pi}\sum_{m=-\infty}^{\infty}e^{im(\phi-\phi')}\int_{0}^{\infty}\mathrm{d}k \ e^{-k\left|z-z'\right|}
     J_m(k\rho)J_m(k\rho'),
\end{equation}
allowing the vector potential to be expressed in terms of cylindrical harmonics defined on a circular plane. Using \eqref{eq.arho}-\eqref{eq.az}, \eqref{eq.streamyd}, and \eqref{green1} we cast the vector potential in a simplified form
\begin{equation}\label{eq.arhon}
    A_{\rho}(\mathbf{r})=\frac{i\mu_0}{2\rho}  \sum_{m=-\infty}^{\infty}e^{im\phi}\int_{0}^{\infty} \mathrm{d}k \ me^{-k|z-z'|}J_m(k\rho)\varphi^m(k),
\end{equation}
\begin{equation}\label{eq.aphin}
    A_{\phi}(\mathbf{r})=-\frac{\mu_0}{2} \sum_{m=-\infty}^{\infty}e^{im\phi}\int_{0}^{\infty} \mathrm{d}k \ ke^{-k|z-z'|}J'_m(k\rho)\varphi^m(k),
\end{equation}
\begin{equation}
    A_z(\mathbf{r})=0,
\end{equation}
where $J'_m(z)$ is the derivative of $J_m(z)$ with respect to $z$, and $\varphi^m(k)$ is defined as the $m^{\textnormal{th}}$ order Hankel transform
\begin{equation}\label{eq.ht}
    \varphi^m(k)=\frac{1}{2\pi}\int_0^\infty{d}\rho' \int_0^{2\pi}\mathrm \mathrm{d}\phi' \ e^{-im\phi'}\rho'J_m(k\rho')\varphi(\rho',\phi).
\end{equation}
We now consider the vector potential generated by the magnetic shield. To do this, we first introduce a pseudo-current density induced on an infinite cylinder, $\tilde{\mathbf{j}}=\tilde{j_\phi}(\phi',z')~\boldsymbol{\hat{\phi}}+\tilde{j_z}(\phi',z')~\mathbf{\hat{z}}$. Next, we seek a Fourier representation of this pseudo-current density, which satisfies the boundary condition over the entire domain of the shield. In particular, we wish to equate the shared azimuthal Fourier modes at the radial boundary of the shield cylinder. This is achieved using a combination of methods that must be applied sequentially, since each method satisfies the condition at an orthogonal boundary. The radial condition is satisfied by equating the magnetic field generated by the cylindrical pseudo-current density and planar current flow, generating a relation between the response of an infinite cylindrical shield and the initial current source. Then, the boundary condition at the end caps can simultaneously be satisfied by applying the method of mirror images~\cite{jackson}. These methods can be combined because the infinite pseudo-current density and planar current flow are spatially orthogonal to the end caps, meaning that any reflections generated by the application of the method of mirror images continue to satisfy the radial condition. The components of the vector potential generated by a pseudo-current density induced on an infinite cylinder~\cite{TurnerBowley1986} in the region $\rho<\rho_s$ are given by
\begin{align}\nonumber
    A_{\rho}\left(\rho,\phi,z\right)=-\frac{i\mu_0\rho_s}{4\pi}\sum_{m=-\infty}^{\infty}\int_{-\infty}^{\infty}\mathrm{d}k \ e^{im\phi}e^{ikz}
     \Big[I_{m-1}(|k|\rho)K_{m-1}(|k|\rho_s) \qquad\qquad\qquad\qquad\qquad\qquad \\ \label{eq.arn} -I_{m+1}(|k|\rho)K_{m+1}(|k|\rho_s)\Big]\tilde{j_\phi}^m(k),
\end{align}
\begin{align}\nonumber
    A_{\phi}\left(\rho,\phi,z\right)=\frac{\mu_0\rho_s}{4\pi}\sum_{m=-\infty}^{\infty}\int_{-\infty}^{\infty}\mathrm{d}k \ e^{im\phi}e^{ikz}
     \Big[I_{m-1}(|k|\rho)K_{m-1}(|k|\rho_s) \ \ \ \ \qquad\qquad\qquad\qquad\qquad\qquad \\ \label{eq.apn} +I_{m+1}(|k|\rho)K_{m+1}(|k|\rho_s)\Big]\tilde{j_\phi}^m(k),
\end{align}
\begin{align} \label{eq.azn}
    A_{z}\left(\rho,\phi,z\right)=\frac{\mu_0\rho_s}{2\pi}\sum_{m=-\infty}^{\infty}\int_{-\infty}^{\infty}\mathrm{d}k \ e^{im\phi}e^{ikz}
     I_m(|k|\rho)K_{m}(|k|\rho_s)\tilde{j_z}^m(k),
\end{align}
where the Fourier transforms of the pseudo-currents are defined by
\begin{equation}\label{eq.fp1}
    j_\phi^m(k)=\frac{1}{2\pi}\int_{0}^{2\pi}\mathrm{d}\phi' \ e^{-im\phi'}\int_{-\infty}^{\infty}\mathrm{d}z'\ e^{-ikz'}\tilde{j_\phi}\left(\phi',z'\right), 
\end{equation}
\begin{equation}\label{eq.fz1}
    j_z^m(k)=\frac{1}{2\pi}\int_{0}^{2\pi}\mathrm{d}\phi' \ e^{-im\phi'}\int_{-\infty}^{\infty}\mathrm{d}z'\ e^{-ikz'}\tilde{j_z}\left(\phi',z'\right).
\end{equation}
The corresponding inverse transforms are given by
\begin{equation}\label{eq.ifp}
    \tilde{j_\phi}(\phi',z')=\frac{1}{2\pi}\sum_{m=-\infty}^{\infty} \int_{-\infty}^{\infty}\mathrm{d}k\ e^{im\phi'}e^{ikz'}j_\phi^m\left(k\right),
\end{equation}
\begin{equation} \label{eq.ifz}
    \tilde{j_z}(\phi',z')=\frac{1}{2\pi}\sum_{m=-\infty}^{\infty} \int_{-\infty}^{\infty}\mathrm{d}k\ e^{im\phi'}e^{ikz'}j_z^m\left(k\right). 
\end{equation}
Therefore, adding the contributions from the planar current flow, \eqref{eq.arhon}-\eqref{eq.aphin}, and the infinite pseudo-current density, \eqref{eq.arn}-\eqref{eq.azn}, while using \eqref{eq.fp1}-\eqref{eq.ifz}, and applying the method of mirror images for two infinitely large parallel planes, we can write the total magnetic field generated by the system in the region $\rho<\rho_s$ as 
\begin{align} \label{eq.shieldyBr}
    B_{\rho}\left(\rho,\phi,z\right)=-\frac{\mu_0}{2\pi}\sum_{p=-\infty}^{\infty}\sum_{m=-\infty}^{\infty}e^{im\phi}\Bigg[\int_{0}^{\infty}\mathrm{d}k \ k^2\frac{(z-(-1)^pz'+pL_s)}{|z-(-1)^pz'+pL_s|}e^{-k|z-(-1)^pz'+pL_s|}J'_m(k\rho)\varphi^m(k) \nonumber \\  -\frac{i\rho_s}{\pi}\int_{-\infty}^{\infty}\mathrm{d}k \ ke^{ikz}
     I'_{m}(|k|\rho)K'_{m}(|k|\rho_s)j_\phi^{mp}(k)\Bigg],
\end{align}
\begin{align} \label{eq.shieldyBp}
    \hspace{-5pt}B_{\phi}\left(\rho,\phi,z\right)=-\frac{i\mu_0}{2\rho}\sum_{p=-\infty}^{\infty}\sum_{m=-\infty}^{\infty}me^{im\phi}\Bigg[\int_{0}^{\infty} \mathrm{d}k \ k\frac{(z-(-1)^pz'+pL_s)}{|z-(-1)^pz'+pL_s|}e^{-k|z-(-1)^pz'+pL_s|}J_m(k\rho)\varphi^m(k)  \nonumber \\ -\frac{i\rho_s}{\pi}\int_{-\infty}^{\infty}\mathrm{d}k \ \frac{|k|}{k}e^{ikz}
     I_{m}(|k|\rho)K'_{m}(|k|\rho_s)j_\phi^{mp}(k)\Bigg],
\end{align}
\begin{align} \label{eq.shieldyBz}
    B_{z}\left(\rho,\phi,z\right)=\frac{\mu_0}{2}\sum_{p=-\infty}^{\infty}\sum_{m=-\infty}^{\infty}e^{im\phi}\Bigg[\int_{0}^{\infty} \mathrm{d}k \ k^2e^{-k|z-(-1)^pz'+pL_s|}J_m(k\rho)\varphi^m(k) \qquad\qquad\qquad\qquad\qquad \nonumber \\ -\frac{\rho_s}{\pi}\int_{-\infty}^{\infty}\mathrm{d}k \ |k|e^{ikz}
     I_{m}(|k|\rho)K'_{m}(|k|\rho_s)j_\phi^{mp}(k)\Bigg],
\end{align}
where $j_\phi^{mp}(k)$ is the $p^{th}$ reflected Fourier transformed azimuthal pseudo-current density induced on the cylindrical surface of the magnetic shield with $I'_m(z)$ and $K'_m(z)$ defined as the derivatives with respect to $z$ of the modified Bessel functions of the first and second kind $I_m(z)$ and $K_m(z)$, respectively. By applying the boundary condition at the radial surface, \eqref{bbound}, we can match the shared $m^{\textnormal{th}}$ azimuthal Fourier mode generated by each $p^\textnormal{th}$ reflected pseudo-current and streamfunction, resulting in the relation
\begin{equation}
    \int_{-\infty}^\infty \mathrm{d}k \ |k|e^{ikz}I_m\left(|k|\rho_s\right)K_m'\left(|k|\rho_s\right)j_\phi^{mp}(k)=\frac{\pi }{\rho_s}\int_{0}^\infty \mathrm{d}k \ k^2 e^{-k|z-(-1)^pz'+pL_s|}J_m\left(k\rho_s\right)\varphi^m(k).
\end{equation}
Physically, due to the formulation of the response in terms of a pseudo-current density, there must be a unique solution that is independent of the axial position that satisfies the boundary condition over the infinite domain of the cylindrical shield. Therefore, we perform an inverse Fourier transform with respect to $z$ to generate an integral representation of the $p^{\textnormal{th}}$ reflected Fourier pseudo-current density, $j_\phi^{mp}(k)$, in terms of the $m^{\textnormal{th}}$ order Hankel transform of the streamfunction defined on the planar surface, $\varphi^m(k)$,
\begin{equation}\label{eq.pseudo}
    j_\phi^{mp}(k)=\frac{e^{-ik\left((-1)^pz'+pL_s\right)}}{\rho_s|k|I_m\left(|k|\rho_s\right)K'_m\left(|k|\rho_s\right)}\int_0^\infty \mathrm{d}\tilde{k} \ \frac{\tilde{k}^3}{\tilde{k}^2+k^2} J_m(\tilde{k}\rho_s) \varphi^m(\tilde{k}).
\end{equation}
This expression for $j_\phi^{mp}(k)$ can now be substituted into \eqref{eq.shieldyBr}-\eqref{eq.shieldyBz} to determine the total magnetic field in terms of $\varphi^m(k)$. Next, we must choose an appropriate expansion of the streamfunction, $\varphi(\rho',\phi')$. Although the choice of orthogonal basis for the expansion of the streamfunction is somewhat arbitrary, a choice of basis which considers the symmetries between the Hankel transform, coordinate system, and the integral representation of the pseudo-current yields a simpler solution. Here, we choose to decompose the radial component of the planar current flow into a Fourier--Bessel series while using a Fourier series representation of the azimuthal dependence,
\begin{equation}\label{eq.streamy}
    \varphi(\rho',\phi')=\left(H\left(\rho'\right)-H\left(\rho-\rho_c\right)\right)\rho_c\sum_{n=1}^N\sum_{m'=0}^{M}J_{m'}\left(\frac{\rho_{nm'}\rho'}{\rho_c}\right)\left(W_{nm'}\cos(m'\phi')+Q_{nm'}\sin(m'\phi')\right),
\end{equation}
where $\left(W_{nm'},Q_{nm'}\right)$ are Fourier coefficients and $\rho_{nm'}$ is the $n^{\textnormal{th}}$ zero of the $m'^{\textnormal{th}}$ Bessel function of the first kind, $J_{m'}\left(\rho_{nm'}\right)=0$. Therefore, using \eqref{eq.shieldyBr}-\eqref{eq.shieldyBz}, \eqref{eq.pseudo}, and \eqref{eq.streamy}, the total magnetic field generated by an arbitrary current flow on the planar surface inside the closed finite high-permeability cylinder can be written as
\begin{equation}\label{eq.cbrff}
    B_{\rho}\left(\rho,\phi,z\right)=\frac{\mu_0\rho_c^3}{2}  \sum_{n=1}^{N}\sum_{m=0}^{M}\rho_{nm}J'_m(\rho_{nm})\left(W_{nm}\cos\left(m\phi\right)+Q_{nm}\sin\left(m\phi\right)\right)B^{nm}_\rho(\rho,z),
\end{equation}
\begin{equation}\label{eq.cbpff}
    B_{\phi}\left(\rho,\phi,z\right)=\frac{\mu_0\rho_c^3}{2\rho} \sum_{n=1}^{N}\sum_{m=0}^{M}m\rho_{nm}J'_m(\rho_{nm})\left(W_{nm}\sin\left(m\phi\right)-Q_{nm}\cos\left(m\phi\right)\right)B^{nm}_\phi(\rho,z),
\end{equation}
\begin{equation}\label{eq.cbzff}
    B_{z}\left(\rho,\phi,z\right)=\frac{\mu_0\rho_c^3}{2} \sum_{n=1}^N\sum_{m=0}^{M}\rho_{nm}J'_m(\rho_{nm})\left(W_{nm}\cos\left(m\phi\right)+Q_{nm}\sin\left(m\phi\right)\right)B^{nm}_z(\rho,z),
\end{equation}
where
\begin{align}
    B^{nm}_\rho(\rho,z)=
    &-\int_0^\infty \mathrm{dk} \ k^2 \sigma\left(k;z,z',L_s\right) \frac{J'_m(k\rho)J_m(k\rho_c)}{k^2\rho_c^2-\rho_{nm}^2}\nonumber \\[6pt]
    &\hspace{140pt} - \sum_{p=1}^\infty \tilde{p}\left|\tilde{p}\right|
    \lambda_p\left(z,z',L_s\right)
    \frac{I'_m(\left|\tilde{p}\right|\rho)I_m(\left|\tilde{p}\right|\rho_c)K_m(\left|\tilde{p}\right|\rho_s)}{I_m(\left|\tilde{p}\right|\rho_s)(\left|\tilde{p}\right|^2\rho_c^2+\rho_{nm}^2)},
\end{align}
\begin{align}
    B^{nm}_\phi(\rho,z)=
    &\int_0^\infty \mathrm{d}k \ k \sigma\left(k;z,z',L_s\right) \frac{J_m(k\rho)J_m(k\rho_c)}{k^2\rho_c^2-\rho_{nm}^2}\nonumber \\[6pt]
    &\hspace{140pt} + \sum_{p=1}^\infty \tilde{p} \ \lambda_p\left(z,z',L_s\right)
    \frac{I_m(\left|\tilde{p}\right|\rho)I_m(\left|\tilde{p}\right|\rho_c)K_m(\left|\tilde{p}\right|\rho_s)}{I_m(\left|\tilde{p}\right|\rho_s)(\left|\tilde{p}\right|^2\rho_c^2+\rho_{nm}^2)},\label{eq.cbpffcoef3}
\end{align}
\begin{align}\nonumber
    B^{nm}_z(\rho,z)=
    &\int_0^\infty \mathrm{d}k \ k^2 \gamma\left(k;z,z',L_s\right) \frac{J_m(k\rho)J_m(k\rho_c)}{k^2\rho_c^2-\rho_{nm}^2} \\[6pt]
    &\hspace{140pt} - \sum_{p=1}^\infty \tilde{p}^2 \
    \tau_p\left(z,z',L_s\right)
    \frac{I_m(\left|\tilde{p}\right|\rho)I_m(\left|\tilde{p}\right|\rho_c)K_m(\left|\tilde{p}\right|\rho_s)}{I_m(\left|\tilde{p}\right|\rho_s)(\left|\tilde{p}\right|^2\rho_c^2+\rho_{nm}^2)}, \label{eq.cbzffcoef3}
\end{align}
and
\begin{equation} \label{eq.gamma_k}
    \gamma\left(k;z,z',L_s\right) = e^{-k|z-z'|} + \frac{2}{e^{2kL_s}-1}\bigg[e^{kL_s}\cosh\left(k\left(z+z'\right)\right) + \cosh\left(k\left(z-z'\right)\right)\bigg],
\end{equation}
\begin{equation}\label{eq.sig_k}
    \sigma\left(k;z,z',L_s\right) = \frac{(z-z')}{|z-z'|} e^{-k|z-z'|} - \frac{2}{e^{2kL_s}-1}\bigg[e^{kL_s}\sinh\left(k\left(z+z'\right)\right) + \sinh\left(k\left(z-z'\right)\right)\bigg],
\end{equation}
\begin{align}\label{eq.S_q_C_q}
    \lambda_p\left(z,z',L_s\right) &= \frac{2}{L_s}\left(\left(-1\right)^p\sin\left(\tilde{p}(z+z')\right) + \sin\left(\tilde{p}(z-z')\right)\right), \\[8pt]
    \tau_p\left(z,z',L_s\right) &= \frac{2}{L_s}\left(\left(-1\right)^p\cos\left(\tilde{p}(z+z')\right) + \cos\left(\tilde{p}(z-z')\right)\right),
\end{align}
with $\tilde{p}=p\pi/L_s$. A full derivation of these expressions is given in Appendix~A. Solving for the unknown Fourier coefficients, $\left(W_{nm},Q_{nm}\right)$, to generate a desired magnetic field using the system of governing equations \eqref{eq.cbrff}-\eqref{eq.cbzff} is an ill-conditioned problem due to the formulation of the vector potential through the integral representation in \eqref{eq.arho}-\eqref{eq.az}. As in previous work by Carlson \emph{et al.}~\cite{doi:10.1002/mrm.1910260202}, this may be solved by a least squares minimization with the addition of a penalty term that acts as a regularization parameter. The choice of regularization parameter is arbitrary since it exists only to facilitate the inversion. Well-regularized solutions yield more simplistic coil designs at a cost to the field fidelity~\cite{Tikhonov1943}. Here, we choose the regularization parameter to be the power, $P$, dissipated by a circular planar current source of thickness, $t$, and resistivity, $\varrho$,
\begin{equation} \label{eq.power}
    P=\frac{\varrho}{t}\int_{0}^{\rho_c}\mathrm{d}\rho'\rho'\int_{0}^{2\pi}\mathrm{d}\phi'\ |J_{\rho}(\rho',\phi')|^2+|J_\phi(\rho',\phi')|^2.
\end{equation}
Substituting \eqref{eq.streamy} into \eqref{eq.streamyd} and then \eqref{eq.power}, and integrating over the planar surface we find that, for $m=0$,
\begin{equation}\label{eq.powerm0}
 P=\frac{\varrho}{t}\pi\rho_c^2\sum_{n=1}^N W_{n0}^2\rho_{n0}^2 J_1\left(\rho_{n0}\right)^2,
\end{equation} 
and, for $m \in \mathbb{Z}^+$,
\begin{align}\nonumber
    P=&\frac{\varrho}{t}\pi\rho_c^2\sum_{n=1}^N\sum_{m=1}^M\left(W_{nm}^2+Q_{nm}^2\right)
    \left(\frac{\rho_{nm}}{2}\right)^{2m}\frac{1}{m!(m-1)!} \\ 
    &\hspace{35pt} \Bigg[\ _2\tilde{F}_3\left(m,m+\frac{1}{2};m+1,m+1,2 m+1;-\rho_{nm}^2\right)\nonumber \\ 
    &\hspace{70pt} -\frac{\rho_{nm}^2}{2(m+1)^2} \ _3\tilde{F}_4\left(m+\frac{1}{2},m+1,m+1;m,m+2,m+2,2 m+1;-\rho_{nm}^2\right)\Bigg]\label{eq.powermmore},
\end{align}
where $_i\tilde{F}_j$ is the regularized hypergeometric function: see Appendix~B for a full derivation. We can now formulate a cost function for the least squares minimization,
\begin{equation} \label{eq.functional}
\Phi=\sum_k^K \left[\mathbf{B}^{desired}\left(\mathbf{r}_k\right)-\mathbf{B}\left(\mathbf{r}_k\right)\right]^2 + \beta P,
\end{equation}
where $\beta$ is a weighting parameter chosen to adjust the physical constraints of the system. The cost function is minimized using a least squares fitting to calculate the optimal Fourier coefficients to generate the desired magnetic field at $K$ target points. The minimization is achieved by taking the differential of the cost function with respect to the Fourier coefficients,
\begin{equation}\label{eq.mindif}
    \frac{\partial \Phi}{\partial W_{ij}}=0, \qquad \frac{\partial \Phi}{\partial Q_{ij}}=0, \qquad i\geq1, \ j\geq0,
\end{equation}
which enables the optimal Fourier coefficients to be found for any given physical target magnetic field profile through matrix inversion. The inversion process yields the optimal continuous streamfunction defined on the planar surface for a finite number of Fourier coefficients. In the ideal case, the number of Fourier coefficients would be infinite. However, a finite number of terms can provide accurate solutions in well-regularized problems. This number can be approximated by ensuring that the distance between the planar current-carrying surface and the closest target field point is much larger than the smallest spatial frequency. This is due to the decreased contribution of higher-order terms at distances much larger than their spatial frequency. The final objective is to design a coil that generates the desired magnetic field to a specified accuracy. To do this, a discrete approximation of the field profile may be found by contouring the streamfunction at $N_\varphi$ discrete levels. The contours of the streamfunction generate streamlines where wires should be laid to replicate the desired target magnetic field. This is achieved by discretizing the streamfunction into $N_{\varphi}$ contours, where $\varphi_j=\textnormal{min}\ \varphi+(j-1/2)\Delta\varphi, j=1,..., N_{\varphi}$, separated by $\Delta \varphi = \frac{\textnormal{max}\ \varphi-\textnormal{min}\ \varphi}{N_{\varphi}}$, and the total current through each wire is $I=\Delta \varphi$. The number of contours should be maximized, limited only by manufacturing since the accuracy of the theoretical model depends on the quality of the discrete approximation of the streamfunction. The approximation to the streamfunction can be determined by using the elemental Biot--Savart law to calculate the error as $N_{\varphi}$ is increased. In the case that multiple current-carrying planes are designed, the contours should be defined evenly between the global maximum and minimum of the streamfunction across all the planes so that the current through each wire is equal.

\section{Results}
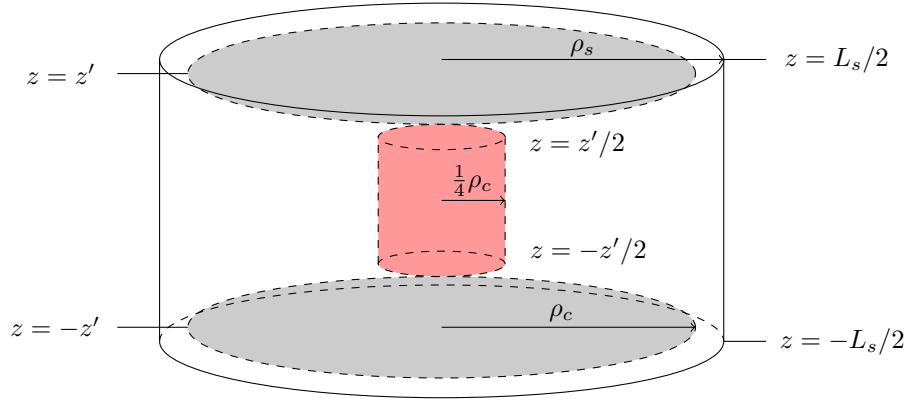
\begin{figure}[htb!]
\begin{center}
\begin{tikzpicture}[scale=0.75]
\draw [black,fill=black!20,dashed] (0,5*0.45) ellipse (10*0.45 and 10*0.09);
\draw [black,fill=black!20,dashed] (0,5*-0.45) ellipse (10*0.45 and 10*0.09);
\draw (0,5*0.5) ellipse (10*0.5 and 10*0.1);
\draw [dashed] (-5,5*-0.5) arc (180:360:5 and -1);
\draw (-5,5*-0.5) arc (180:360:5 and 1);
\draw (-5,2.5) -- (-5,-2.5);
\draw (5,2.5) -- (5,-2.5);
\node at (7,2.5) {$z=L_s/2$};
\node at (7.1,-2.5) {$z=-L_s/2$};
\draw (5,2.5) -- (5.75,2.5);
\draw (5,-2.5) -- (5.75,-2.5);
\node at (-6.75,2.25) {$z=z'$};
\node at (-6.85,-2.25) {$z=-z'$};
\draw (-4.5,2.25) -- (-5.75,2.25);
\draw (-4.5,-2.25) -- (-5.75,-2.25);
\node at (2.4,1) {$z=z'/2$};
\node at (2.6,-0.9) {$z=-z'/2$};
\node at (2.5,2.7) {$\rho_s$};
\node at (2.125,-2.0) {$\rho_c$};
\draw [->] (0,2.5) -- (5,2.5);
\draw [->] (0,-2.25) -- (4.5,-2.25);
\draw[black,fill=red!40,dashed] (-10*0.45/4,-5*0.45/2) rectangle ++(10*0.45/2,5*0.45/1);
\draw [black,fill=red!40,dashed] (0,5*0.45/2) ellipse (10*0.45/4 and 10*0.09/4);
\draw [black,fill=red!40,dashed] (0,5*-0.45/2) ellipse (10*0.45/4 and 10*0.09/4);
\draw [->] (0,0) -- (10*0.45/4,0);
\node at (2.125/4,0.35) {$\frac{1}{4}\rho_c$};
\end{tikzpicture}
\end{center}
\caption{Schematic diagram showing the areas occupied by the bi-planar coils (light gray) inside a cylindrical closed high-permeability magnetic shield (black outline). The high-permeability shield is of length $L_s=1$ m and radius $\rho_s=0.5$ m, with planar end caps located at $z=\pm L_s/2$. The shield encloses interior conducting planes of radius $\rho_c=0.45$ m which are located at $z'=\pm0.45$ m. The optimization region (red, enclosed by hatched lines) is bounded along the $z$-axis between the coil planes with top and bottom positions $z=\pm z'/2$, respectively, and extends radially between $\rho=[0,\rho_c/4]$.}
\label{fig.shieldandcoil}
\end{figure}

We now verify our analytical model by designing hybrid active--passive systems composed of bi-planar coils inside a closed high-permeability magnetic shield and compare the resulting magnetic field profiles with forward numerical simulations of each optimized system. We consider two distinct systems, each containing current confined to two disks of radius $\rho_c=0.45$ m and symmetrically placed at $z'=\pm 0.45$ m. Both systems are interior to a perfect closed cylindrical magnetic shield of radius $\rho_s=0.5$ m and length $L_s=1$ m centered on the origin, as shown in Fig.~\ref{fig.shieldandcoil}. The first system is designed to generate a constant transverse field, $\mathbf{B}=B_0\mathbf{\hat{x}}$, and the second system creates a linear field gradient, $\mathbf{B}=G(-x~\mathbf{\hat{x}}-y~\mathbf{\hat{y}}+2z~\mathbf{\hat{z}})$, within an optimization region defined by $-z'/2\leq z \leq z'/2$ and $0\leq\rho\leq \rho_c/4$. The magnetic field profiles chosen, i.e. the uniform transverse and linear field gradient fields, are examples of tesseral ($m\neq0$) and zonal ($m=0$) harmonics, respectively, which exhibit $m$-fold and complete azimuthal symmetry~\cite{doi:10.1002/mrm.1910010107}, which facilitates analysis of the shield's particular response to tesseral and zonal harmonic fields generated by planar current sources. The two systems that we consider here are chosen to illustrate targeted magnetic field compensation in situations where compact systems are required, but space inside the central cylindrical cavity of the system is limited by the presence of experimental equipment (e.g. magnetic sensors). Coil designs generating other field profiles, for example using combinations of planar coils with varying sizes, can be found in the Appendix~C. All designs can be replicated using our open-access Python code\cite{pjpython}.

\begin{figure}[htb!]
\captionsetup[subfigure]{labelformat=empty}
\begin{center}
     \begin{subfigure}{0.49\textwidth}
         \centering
         \includegraphics[width=\textwidth]{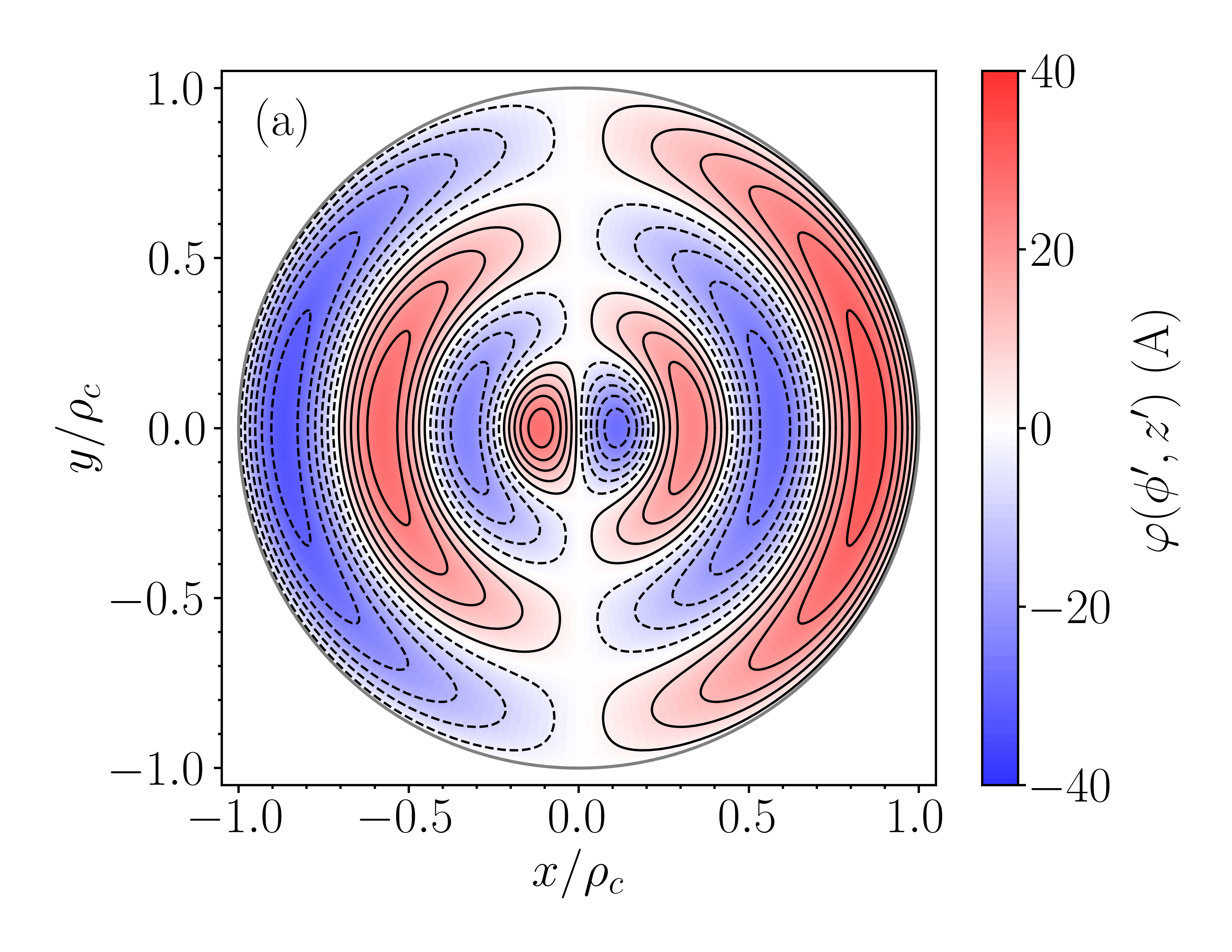}
         \caption{}
         \label{fig.bxtop}
     \end{subfigure}
     \begin{subfigure}{0.49\textwidth}
         \centering
         \includegraphics[width=\textwidth]{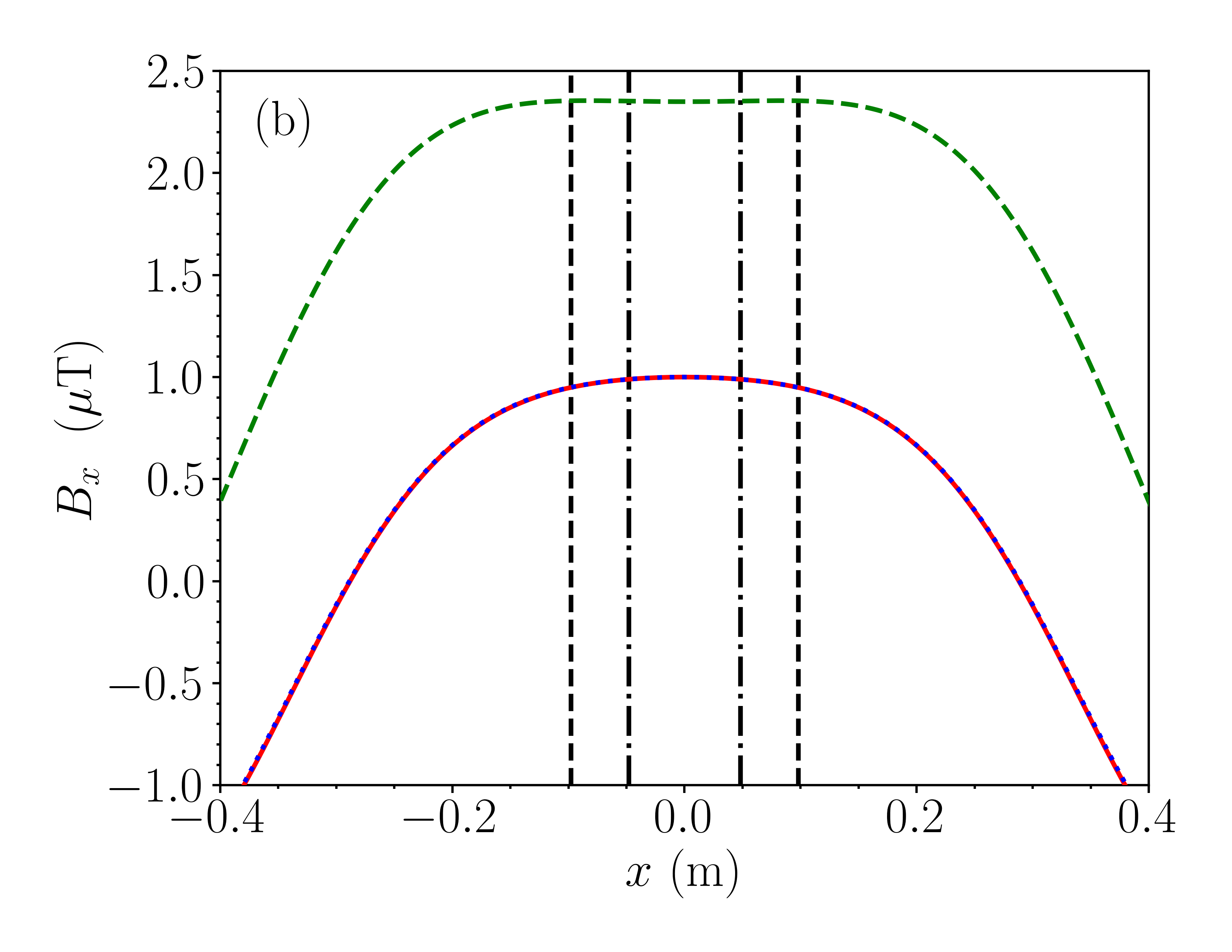}
         \caption{}
         \label{fig.bxradial}
     \end{subfigure}
     \begin{subfigure}{0.98\textwidth}
         \centering
         \includegraphics[width=\textwidth]{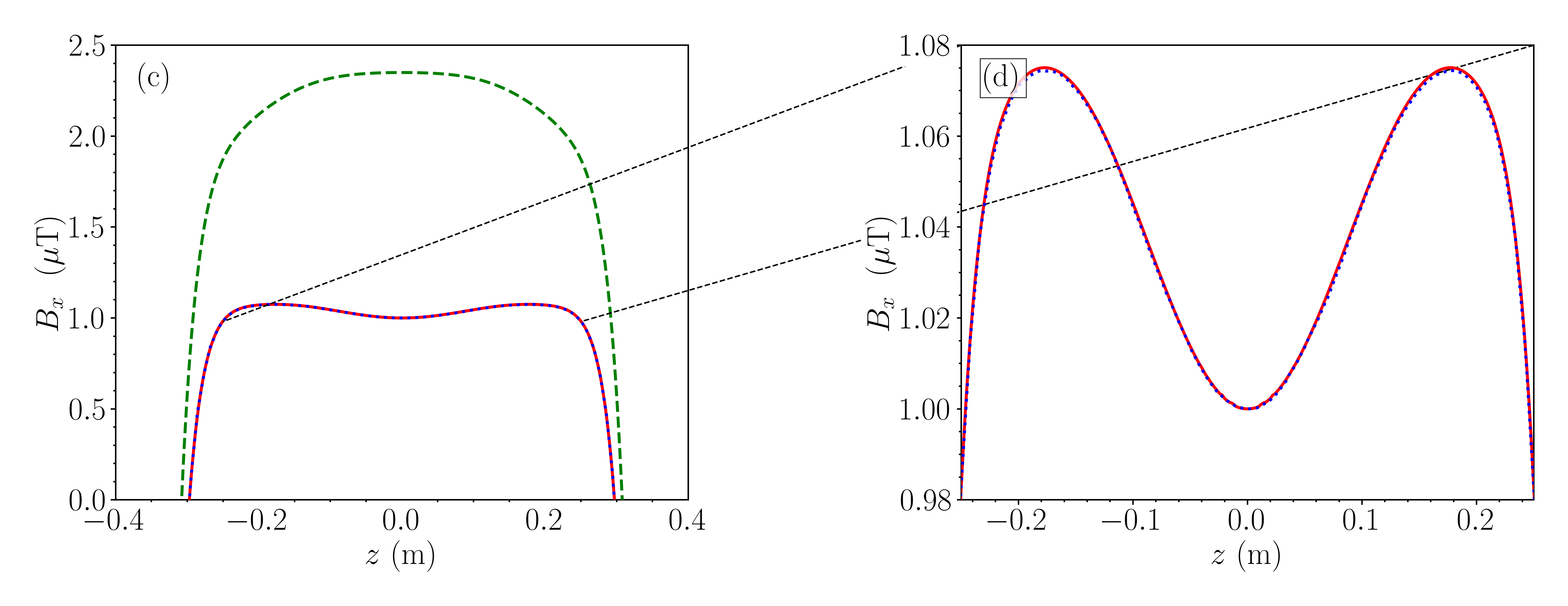}
         \caption{}
         \label{fig.bxaxial}
     \end{subfigure}
\end{center}
\caption{Wire layouts (a) and performance (b--d) of a hybrid active--passive system optimized to generate a constant transverse magnetic field, $B_x$. Current flows on the surface of two disks of radius $\rho_c=0.45$ m, which are separated symmetrically from the origin and lie at $z'=\pm0.45$ m. The wire layouts are optimized to generate a constant transverse field, $B_0=1$~$\mu$T, across the cylinder and normal to its axis of symmetry. The current-carrying planes are placed symmetrically inside a perfect closed magnetic shield of radius $\rho_s=0.5$ m and length $L_s=1$ m and the magnetic field is optimized between $\rho=[0,\rho_c/4]$ and $z=\pm{z'/2}$, as shown in Fig.~\ref{fig.shieldandcoil}. The least squares optimization was performed with parameters $N=50$, $M=1$, $\beta=1.77\times10^{-9}$~T\textsuperscript{2}/W, $t=0.5$~mm, and $\rho=1.68\times10^{-8}$~$\Omega$m. (a) Color map of the optimal current streamfunction calculated for the upper current-carrying plane in Fig.~\ref{fig.shieldandcoil}. Blue and red shaded regions correspond to current counter flows and their intensity shows the streamfunction magnitude from low (white) to high (intense color). Solid and dashed black curves represent discrete wires with opposite senses of current flow, approximating the current continuum with $N_\varphi=12$ global contour levels. Streamfunction on the lower light gray plane in Fig.~\ref{fig.shieldandcoil} is geometrically identical but the current direction is reversed. (b) $B_x$, calculated versus transverse position, $x$, for $y=z=0$, from the current continuum in (a) in three ways: analytically using \eqref{eq.cbrff}-\eqref{eq.cbzff} (solid red curve); numerically using COMSOL Multiphysics\textsuperscript{\textregistered} Version 5.3a, modeling the high-permeability cylinder as a perfect magnetic conductor (blue dotted curve); numerically \emph{without} the high-permeability cylinder and using the Biot–Savart law with $N_\varphi=100$ contour levels (dashed green curve). Black lines enclose the regions where the calculated field deviates from the target field by $5\%$ (dashed) and $1\%$ (dot-dashed). (c) $B_x$, calculated versus axial position, $z$, for $x=y=0$, from the current continuum in the same three ways as for (b). (d) Enlarged section of (c) showing good agreement between the numerical and analytical results throughout the optimization region.}
\label{fig.bxcoil}
\end{figure}

\begin{figure}[htb!]
\captionsetup[subfigure]{labelformat=empty}
\begin{center}
     \begin{subfigure}{0.49\textwidth}
         \centering
         \includegraphics[width=\textwidth]{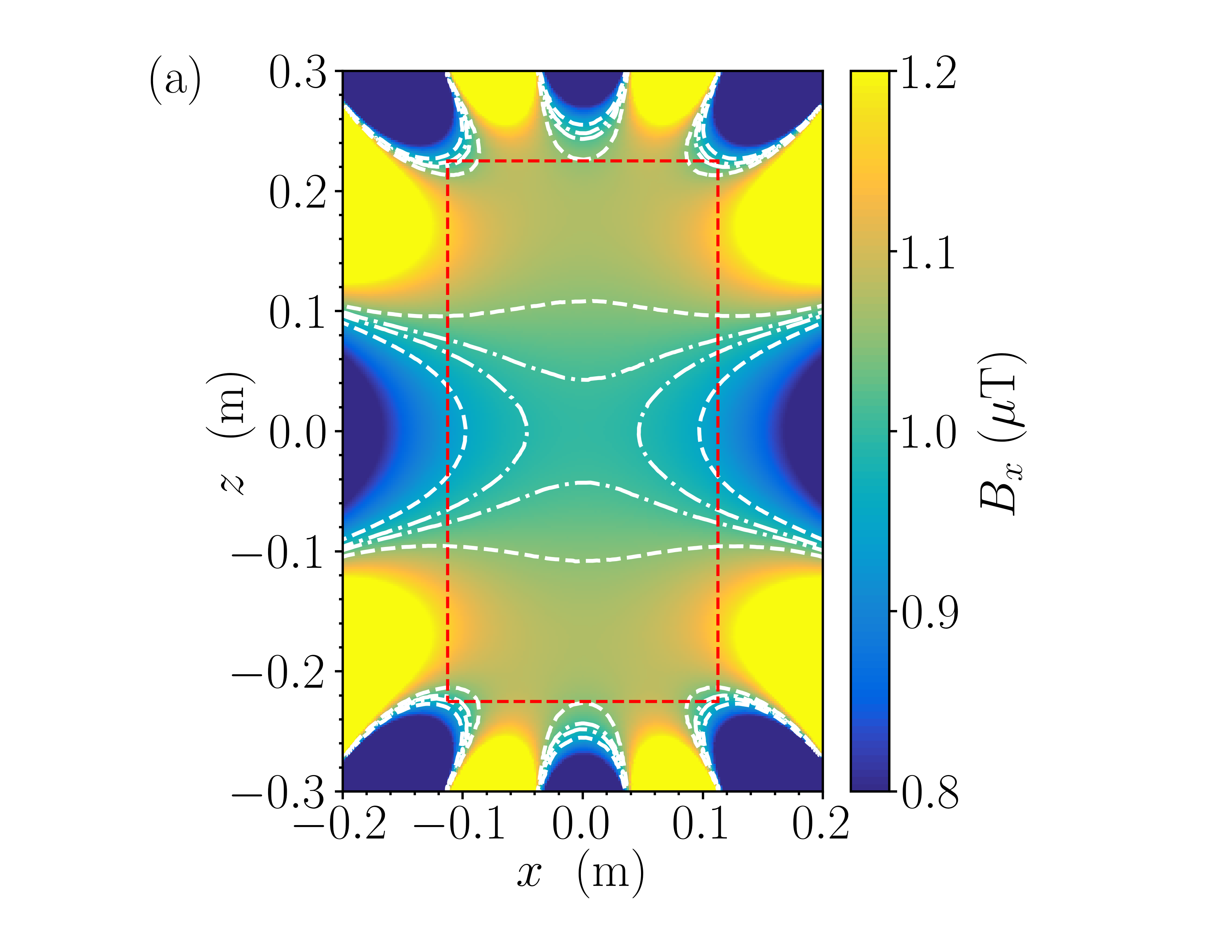}
         \caption{}
         \label{fig.bxcolormapsx}
     \end{subfigure}
     \begin{subfigure}{0.49\textwidth}
         \centering
         \includegraphics[width=\textwidth]{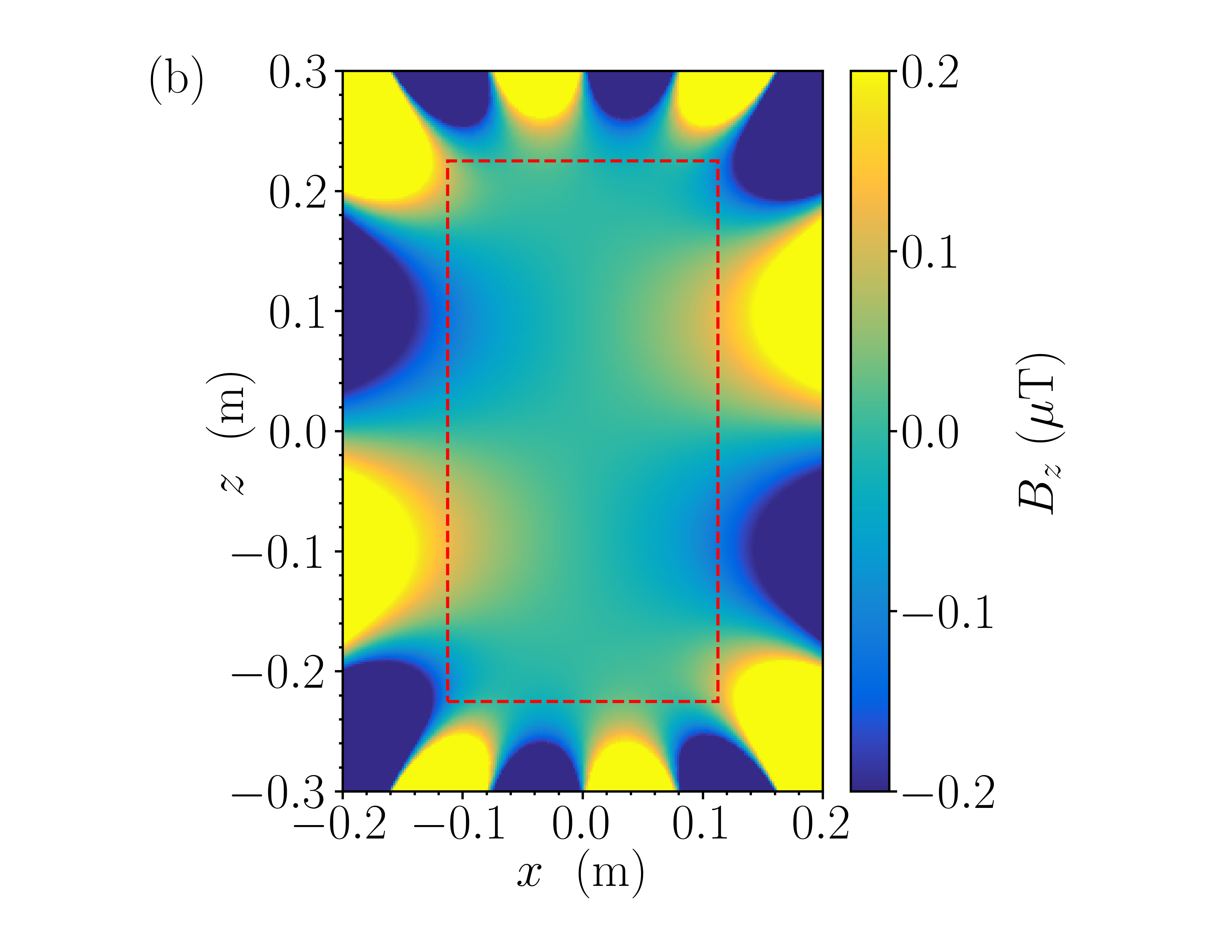}
         \caption{}
         \label{fig.bxcolormapsz}
     \end{subfigure}
\end{center}
\caption{Color maps showing the magnitude of the magnetic field, in the $xz$-plane inside a closed finite length perfect magnetic conductor generated by the active--passive system depicted in Fig.~\ref{fig.bxcoil}: (a) transverse component, $B_x$, and (b) axial component, $B_z$. The field profiles were calculated numerically using COMSOL Multiphysics\textsuperscript{\textregistered} Version 5.3a. The magnetic field is optimized between $\rho=[0,\rho_c/4]$ and $z=\pm{z'/2}$; dashed red lines in (a--b). Contours, in (a) only, show where the field deviates from the target field by $5$\% (dashed curves) and $1$\% (dot-dashed curves).}
\label{fig.bxcolormaps}
\end{figure}

\begin{figure}[htb!]
\captionsetup[subfigure]{labelformat=empty}
\begin{center}
     \begin{subfigure}{0.49\textwidth}
         \centering
         \includegraphics[width=\textwidth]{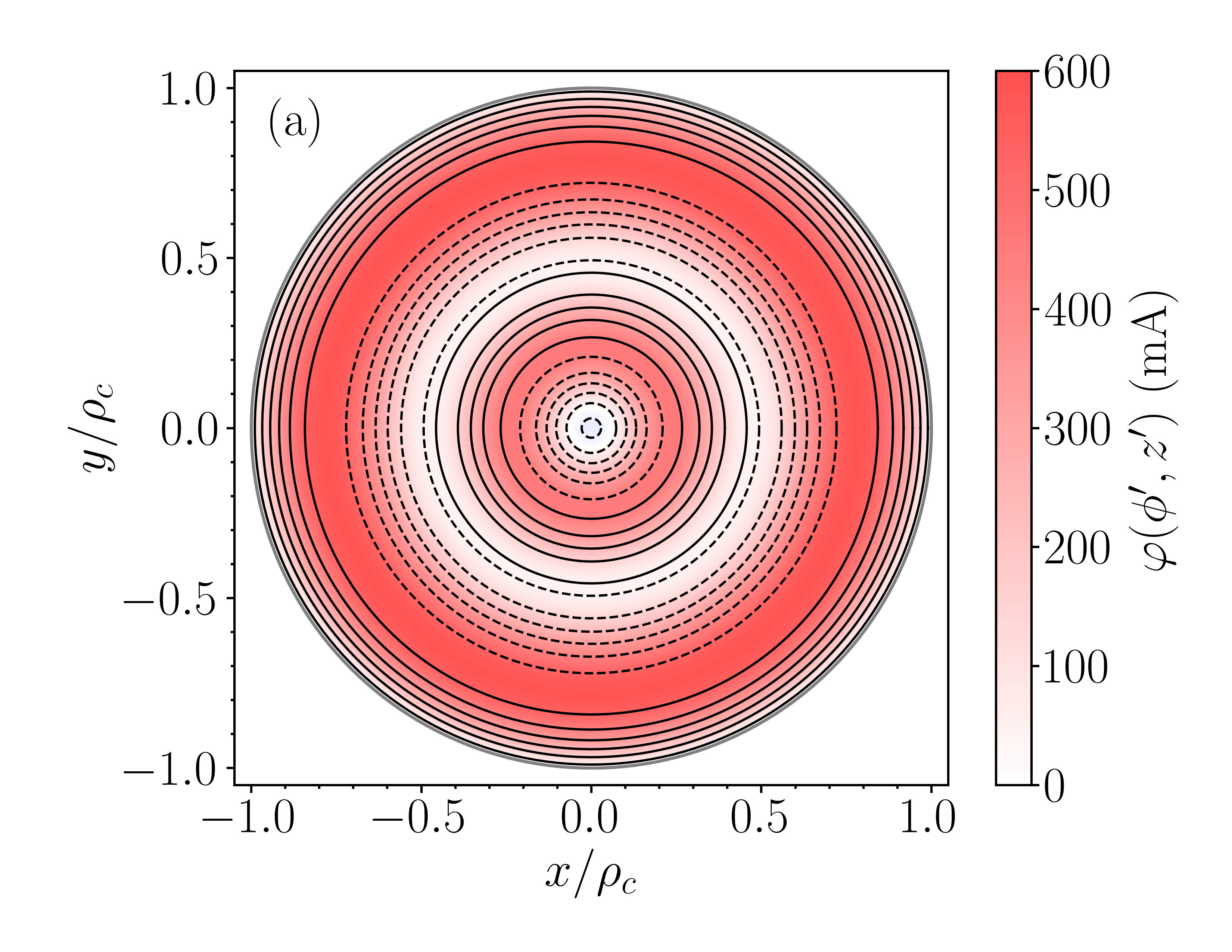}
         \caption{}
         \label{fig.dbztop}
     \end{subfigure}
     \begin{subfigure}{0.49\textwidth}
         \centering
         \includegraphics[width=\textwidth]{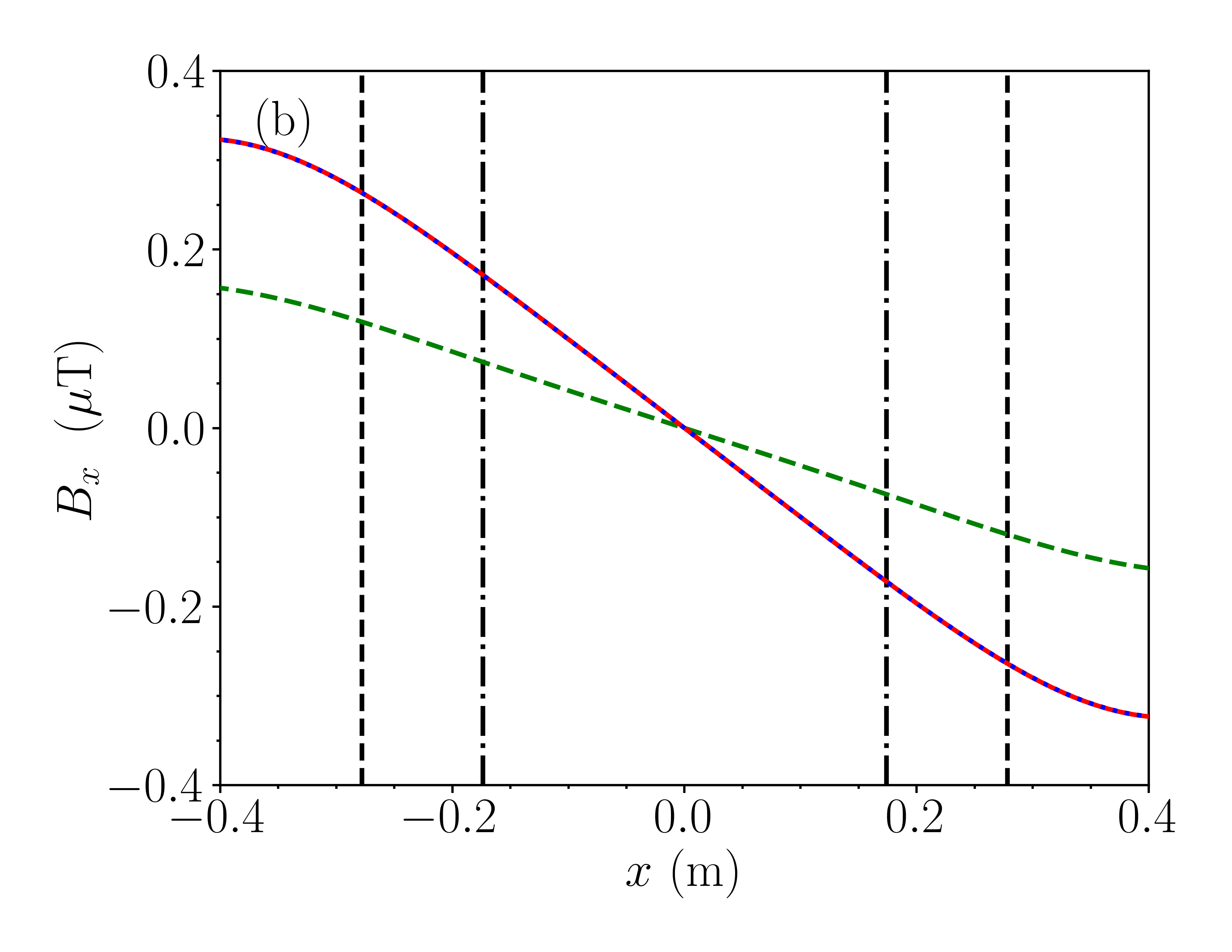}
         \caption{}
         \label{fig.dbzradial}
     \end{subfigure}
     \begin{subfigure}{0.49\textwidth}
         \centering
         \includegraphics[width=\textwidth]{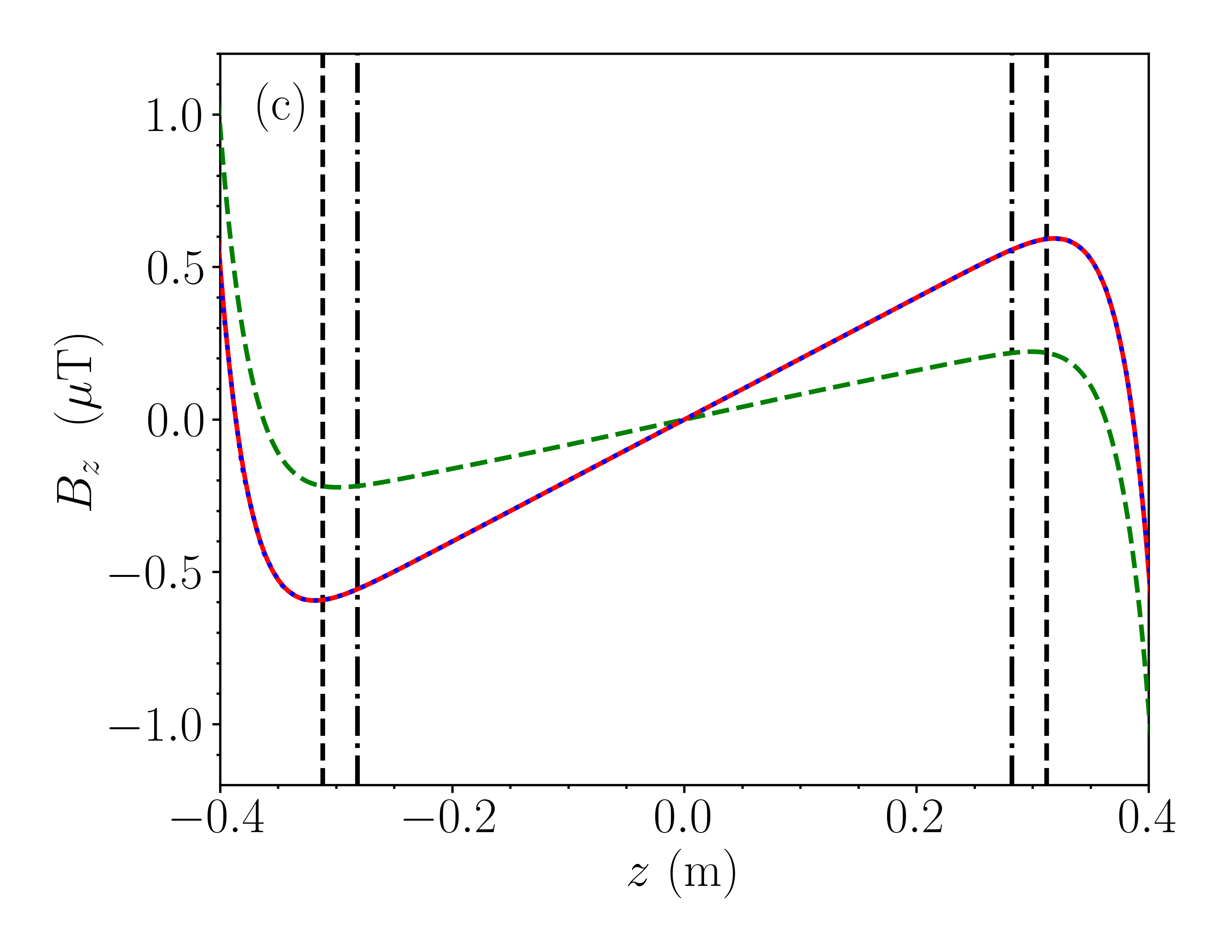}
         \caption{}
         \label{fig.dbzaxial}
     \end{subfigure}
\end{center}
\caption{Wire layouts (a) and performance (b--c) of a hybrid active--passive system optimized to generate a linear variation of $B_z$ with $z$-position. Current flows on the surface of two disks of radius $\rho_c=0.45$ m, which are separated symmetrically from the origin and lie at $z'=\pm0.45$ m. The wire layouts are optimized to generate a linear axial field gradient, $\mathrm{d}B_z/\mathrm{d}z=2$ $\mu$T/m, along the $z$-axis of the cylinder through the harmonic $\mathbf{B}=G(-x~\mathbf{\hat{x}}-y~\mathbf{\hat{y}}+2z~\mathbf{\hat{z}})$. The current-carrying planes are placed symmetrically inside a perfect closed magnetic shield of radius $\rho_s=0.5$ m and length $L_s=1$ m and the magnetic field is optimized between $\rho=[0,\rho_c/4]$ and $z=\pm{z'/2}$, as shown in Fig.~\ref{fig.shieldandcoil}. The least squares optimization was performed with parameters $N=50$, $M=0$, $\beta=1.77\times10^{-9}$~T\textsuperscript{2}/W, $t=0.5$~mm, and $\rho=1.68\times10^{-8}$~$\Omega$m. (a) Color map of the optimal current streamfunction calculated for the upper current-carrying plane in Fig.~\ref{fig.shieldandcoil}. Intensity of red shaded regions shows the streamfunction magnitude from low (white) to high (intense color). Solid and dashed black curves represent discrete wires with opposite senses of current flow, approximating the current continuum with $N_\varphi=12$ global contour levels. Streamfunction on the lower light gray plane in Fig.~\ref{fig.shieldandcoil} is geometrically identical but the current direction is reversed.  (b) Transverse magnetic field, $B_x$, calculated versus transverse position, $x$, for $y=z=0$, from the current continuum in (a) in three ways: analytically using \eqref{eq.cbrff}-\eqref{eq.cbzff} (solid red curve); numerically using COMSOL Multiphysics\textsuperscript{\textregistered} Version 5.3a, modelling the high-permeability cylinder as a perfect magnetic conductor (blue dotted curve); numerically \emph{without} the high-permeability cylinder and using the Biot–Savart law with $N_\varphi=100$ contour levels (dashed green curve). Black lines enclose the regions where the calculated field deviates from the target field by $5\%$ (dashed) and $1\%$ (dot-dashed). (c) Axial magnetic field, $B_z$, calculated versus axial position, $z$, for $x=y=0$, from the current continuum in (a) in the same three ways as for (b).}
\label{fig.bzcoil}
\end{figure}

In Fig.~\ref{fig.bxcoil}a and Fig.~\ref{fig.bzcoil}a, we show the optimized contoured streamfunctions for the constant transverse and linear field gradient systems, respectively. Due to the symmetric placement of the bi-planar coil systems and optimized field regions within the shield (see Fig.~\ref{fig.shieldandcoil}), the magnitudes of the streamfunctions defined on both coils are identical. The current flow directions, however, are opposite, due to the form of the desired fields. Figures.~\ref{fig.bxcoil}b--d show the transverse magnetic field, $B_0=1$~$\mu$T, along the $x$-axis and $z$-axis, respectively, generated by the bi-planar coil design shown in Fig.~\ref{fig.bxcoil}a calculated in three different ways: analytically using \eqref{eq.cbrff}-\eqref{eq.cbzff} (solid red curves); numerically using COMSOL Multiphysics\textsuperscript{\textregistered} with the shield treated as a perfect magnetic conductor (blue dotted curves); numerically in free space, i.e. excluding the high-permeability material and evaluating the magnetic fields through the Biot--Savart law for discretized bi-planar coils with $N_\varphi=250$ (dashed green curve in b). Furthermore, color maps of the transverse and axial magnetic field components in the $xz$-plane are presented in Fig.~\ref{fig.bxcolormapsx} and Fig.~\ref{fig.bxcolormapsz}, respectively, calculated numerically using COMSOL Multiphysics\textsuperscript{\textregistered} with the shield treated as a perfect magnetic conductor. Similarly, Fig.~\ref{fig.bzcoil}b--c show the linear axial field gradient, $G=1$~$\mu$T/m, along the $x$-axis and $z$-axis, respectively, generated by the bi-planar coil system shown in Fig.~\ref{fig.bzcoil}a in the same three cases with similar color maps of the magnetic field components in the $xz$-plane shown in Fig.~\ref{fig.bzcolormapsx} and Fig.~\ref{fig.bzcolormapsz}. For both designs, the analytical field profiles agree well with numerical simulations. The maximum errors between the analytical model and numerical simulation are $0.052$\% and $0.043$\%, for the constant transverse field and the gradient of the linear gradient field, respectively, along the $z$-axis of the optimization region.

To quantify the performance of our optimization procedure, we can analyze the deviations between the magnetic fields generated by our theoretical model and the desired target fields. Examining the fidelity of the fields generated by both systems along $x$-axis of the optimization region, the maximum absolute deviations from the constant transverse and linear axial gradient fields are $6.78\%$ and $0.380\%$, respectively. Along the $z$-axis of the optimization region, the maximum deviations are $7.50\%$ and $0.306\%$, respectively. Clearly, the axial field gradient field is generated more accurately than the uniform transverse field. This can be seen from Fig.~\ref{fig.bxcoil}d, which reveals small oscillations in the transverse field over the $z$-axis of the optimization region. To understand the difference between the fidelity of the field profiles generated by the two systems, we must analyze how the passive magnetic shield affects the fields from the planar current distributions, decomposing its response into zonal ($m=0$) and tesseral ($m\neq 0$) harmonic components. These harmonic responses relate to the variations in the induced pseudo-current density, corresponding to the planar streamfunction \eqref{eq.pseudo}, required to satisfy the boundary condition at the shield wall. Schematic approximations of the surface currents for the $m=0$ zonal and $m=1$ tesseral harmonic responses can be seen in Fig.~\ref{fig.responsediagram}. For the boundary condition on the cylindrical surface to be satisfied, the induced azimuthal pseudo-currents must mirror azimuthal current paths on the planar coil surfaces, so that the associated fields cancel in the region $\rho_s>\rho>\rho_c$. Since zonal responses are composed of simple circular loops, which can be formed in either a cylindrical or planar basis, the response of the magnetic shield enhances the magnetic field in the optimization region. Consequently, the shield amplifies the axial magnetic field by a factor of $2.39$ at the shield's center, as shown in Fig.~\ref{fig.bzcoil}b--c. The uniform field gradient therefore exhibits superior fidelity because the continuum response of the passive shield approximates a distributed cylindrical coil. The resulting system, composed of both the coil and shield, completely encloses the interior region, producing a high-fidelity magnetic field gradient.
The tesseral responses are more complicated, with the cylindrical surface of the magnetic shield acting to oppose the magnetic field generated by the planar system in the region $\rho<\rho_s$ due to the formation of saddle-type currents~\cite{doi:10.1002/mrm.1910010107}. These currents result from the required continuity of the induced azimuthal pseudo-current density in a cylindrical basis. Due to the restricted current flow on the planar surface, the only way to mitigate the opposing field generated by the cylindrical surface of the shield is to minimize the magnetic field at the boundary, resulting in field coil designs that are oscillatory, as seen in Fig.~\ref{fig.bxcoil}a. The response of the passive shield, for this configuration, reduces the magnetic field by a factor of $2.41$ at the shield's center, as shown in Fig.~\ref{fig.bxcoil}b--c.

\begin{figure}[htb!]
\captionsetup[subfigure]{labelformat=empty}
\begin{center}
     \begin{subfigure}{0.49\textwidth}
         \centering
         \includegraphics[width=\textwidth]{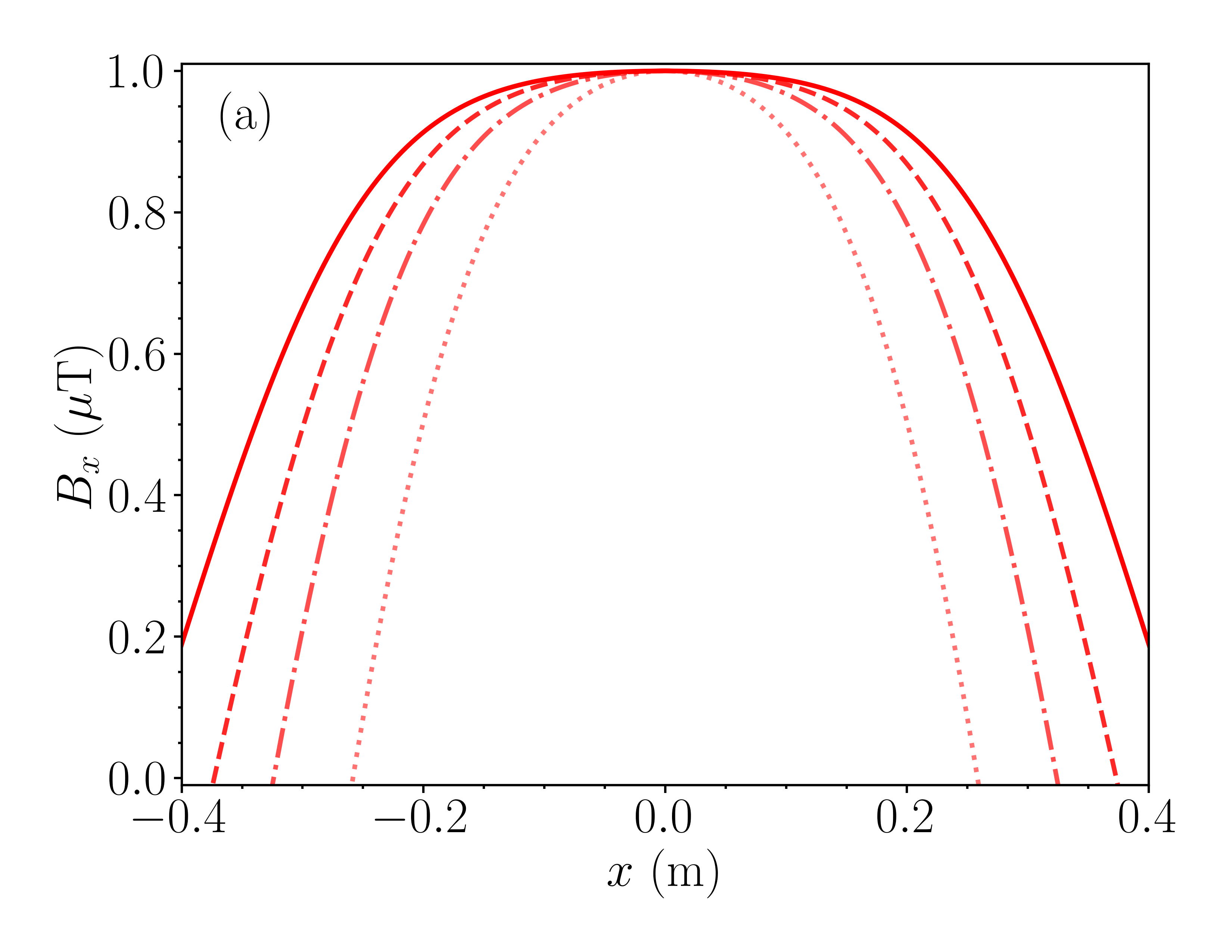}
         \caption{}
         \label{fig.bxprofs}
     \end{subfigure}
     \begin{subfigure}{0.49\textwidth}
         \centering
         \includegraphics[width=\textwidth]{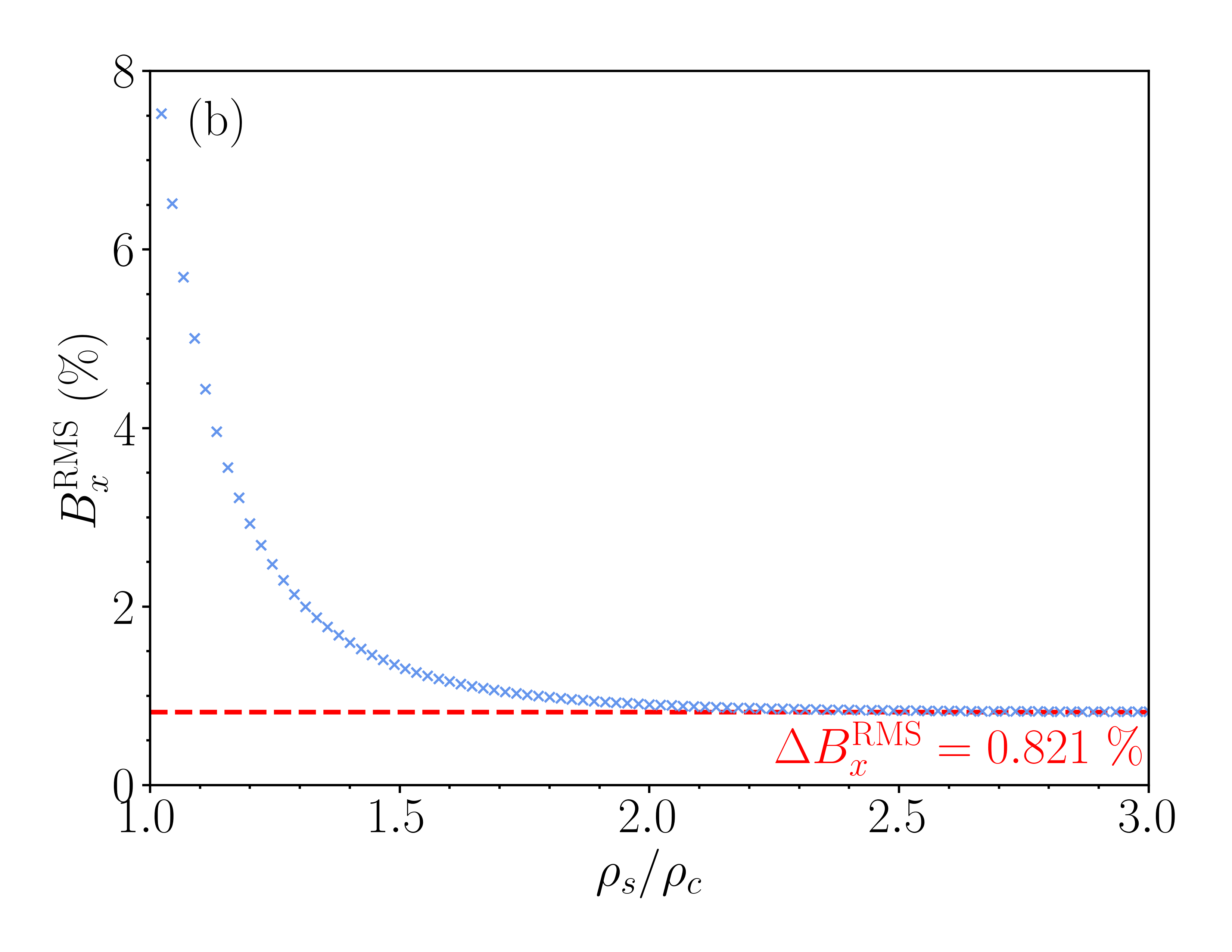}
         \caption{}
         \label{fig.bxscan}
     \end{subfigure}
\end{center}
\caption{Improvement in the performance of a hybrid active--passive system, optimized to generate a constant transverse field $B_0=1$ $\mu$T, upon increasing the radius of the passive shield. Current flows on the surface of two disks of radius $\rho_c=0.45$ m, which are separated symmetrically from the origin and lie at $z'=\pm0.45$ m. The wire layouts are optimized to generate constant $B_0=1$ $\mu$T between $\rho=[0,\rho_c/4]$ and $z=\pm{z'/2}$, as shown in Fig.~\ref{fig.shieldandcoil}. The current-carrying planes lie within the bore of a perfect closed magnetic shield of radius $\rho_s=[\rho_c,3\rho_c]$ and length $L_s=1$ m. Each least squares optimization was performed with parameters $N=50$, $M=1$, $\beta=1.77\times10^{-9}$~T\textsuperscript{2}/W, $t=0.5$~mm, and $\rho=1.68\times10^{-8}$~$\Omega$m. (a) Transverse magnetic field, $B_x$, calculated analytically versus transverse position, $x$, using \eqref{eq.cbrff}-\eqref{eq.cbzff} (red curves), where: $\rho_s/\rho_c=1.01$ (dotted curve), $\rho_s/\rho_c=1.25$ (dash-dotted curve), $\rho_s/\rho_c=1.5$ (dashed curve), and $\rho_s/\rho_c=2.5$ (solid curve). (b) Root mean square (RMS) deviation, ${\Delta}B_x^\mathrm{RMS}$, between the calculated and target fields evaluated along the axis of the optimized field region and plotted as a function of $\rho_s/\rho_c$. Light blue crosses show ${\Delta}B_x^\mathrm{RMS}$ values calculated analytically using \eqref{eq.cbrff}-\eqref{eq.cbzff}. Horizontal dashed red line shows the analytical value of ${\Delta}B_x^\mathrm{RMS} = 0.821 \%$ obtained in the wide shield limit ($\rho_s/\rho_c>>1$).}
\label{fig.bxiterplot}
\end{figure}

The shield's response not only makes the optimization of tesseral harmonic fields more difficult in regions close to the cylindrical surface of the magnetic shield, but also has an effect on the level of fidelity that can be achieved over any region when the coils are in close proximity to the magnetic shield. To demonstrate this, we investigate the fidelity of the constant transverse field generated by optimized bi-planar coils, similar to those in Fig.~\ref{fig.bxcoil}, within a magnetic shield whose radius varies over the range $\rho_s=[\rho_c,3\rho_c]$. In Fig.~\ref{fig.bxiterplot}a we plot field profiles at selected shield radii to show how the uniformity of the transverse field along the $x$-axis improves as the radius of the shield increases, as expected, due to the reduced inhomogeneities introduced by the cylindrical surface of the magnetic shield at distances further from the field coils. To evaluate the deviation in the field uniformity, we can calculate the mean RMS error, which represents the averaged deviation of the transverse field from perfect uniformity. In Fig~\ref{fig.bxiterplot}b, we evaluate the mean RMS error in the transverse field along the $z$-axis of the optimization region as the radius of the magnetic shield increases. The uniformity of the optimized fields in Fig.~\ref{fig.bxiterplot}a and Fig.~\ref{fig.bxiterplot}b initially decreases rapidly as the radius of the magnetic shield increases, but approaches the wide shield limit when $\rho_s \sim 2\rho_c$, at which point the field generated by the cylindrical surface of the magnetic shield becomes negligible. In the wide shield limit the predominant contribution from the magnetic shield can then be approximated through the use of the method of mirror images from two infinite parallel perfect magnetic conductors.

Although the pseudo-currents do not exist physically, analysis of the magnetic fields generated by the shield in response to the planar coils gives insight into how such coils should be designed in order to best realize a specified target field. For example, generating magnetic fields that require coil geometries with $m$-fold azimuthal symmetry causes the cylindrical surface of the magnetic shield to oppose the magnetic field generated by the planar coil. Consequently, if a tesseral field is required over a large radial region, the distance from this region to the planar coils and from the cylindrical shield surface should be minimized and maximized, respectively. The opposite is true for the fields generated by the planar end caps of the shield. When the bi-planar coils are located near the end caps, where $-L_s/2<-z'<z<z'<L_s/2$, but radially distant from the cylindrical surface of the magnetic shield, the reflected pseudo-currents generated by the method of mirror images provide a field similar to that of the planar coils, resulting in a field that is magnified. Consequently, bi-planar coils are desirable in magnetic shields with large radii and small aspect ratios, $L_s/\rho_s<1$. In comparison, the magnetic field generated by the shield in response to coils contained on a \emph{cylindrical} surface~\cite{PhysRevApplied.14.054004} shows the opposite effect. Tesseral fields generated by cylindrical coils are enhanced by the cylindrical surface and are reduced by the planar end caps, favoring long magnetic shields with $L_s/\rho_s>1$. Due to the conflicting conditions on the geometries (i.e. planar or cylindrical) of the coils and the magnetic shield, a system composed of coupled planar and cylindrical coils may have advantages for generating desired tesseral field profiles with the greatest accuracy.

Intrinsic inaccuracies in calculating the optimum current density and approximating the current continuum by discrete wires are not the only sources of error that must be considered when designing fields using our method. Particularly for the tesseral harmonic field generating coils, wire patterns may be highly meandering, making them hard to manufacture and, due to their high resistance, power consuming. When manufacturing these systems, the achievable field fidelity is limited by the discretization of the continuum current flow pattern and by the creation of magnetic dipoles via the interaction between separate current streamlines and the wires that follow them. Fortunately, new ways to realize the current continuum are emerging due to advances in foldable PCBs~\cite{PCBCoils} and 3D-printing technologies~\cite{3DPrintingPaper}, which enable more complex coils to be made accurately. It should also be noted that approximating a perfect magnetic shield by a material of finite permeability $\mu_r=20000$ and thickness $d=1$ mm only introduces small deviations in the model of $0.005$\%~\cite{PhysRevApplied.14.054004}, which is much less than the error introduced in the desired field by the coil designs. As a result, given an accurate representation of the current continuum, our methodology can be used reliably to generate target magnetic fields in high-permeability environments. The python code used to design arbitrary the planar coils in a magnetically shielded cylinder is openly available from GitHub and can be cited at http://doi.org/10.5281/zenodo.4442661 \cite{pjpython}. Verification using COMSOL Multiphysics\textsuperscript{\textregistered} requires a valid license.
\begin{figure}[htb!]
\captionsetup[subfigure]{labelformat=empty}
\begin{center}
     \begin{subfigure}{0.49\textwidth}
         \centering
         \includegraphics[width=\textwidth]{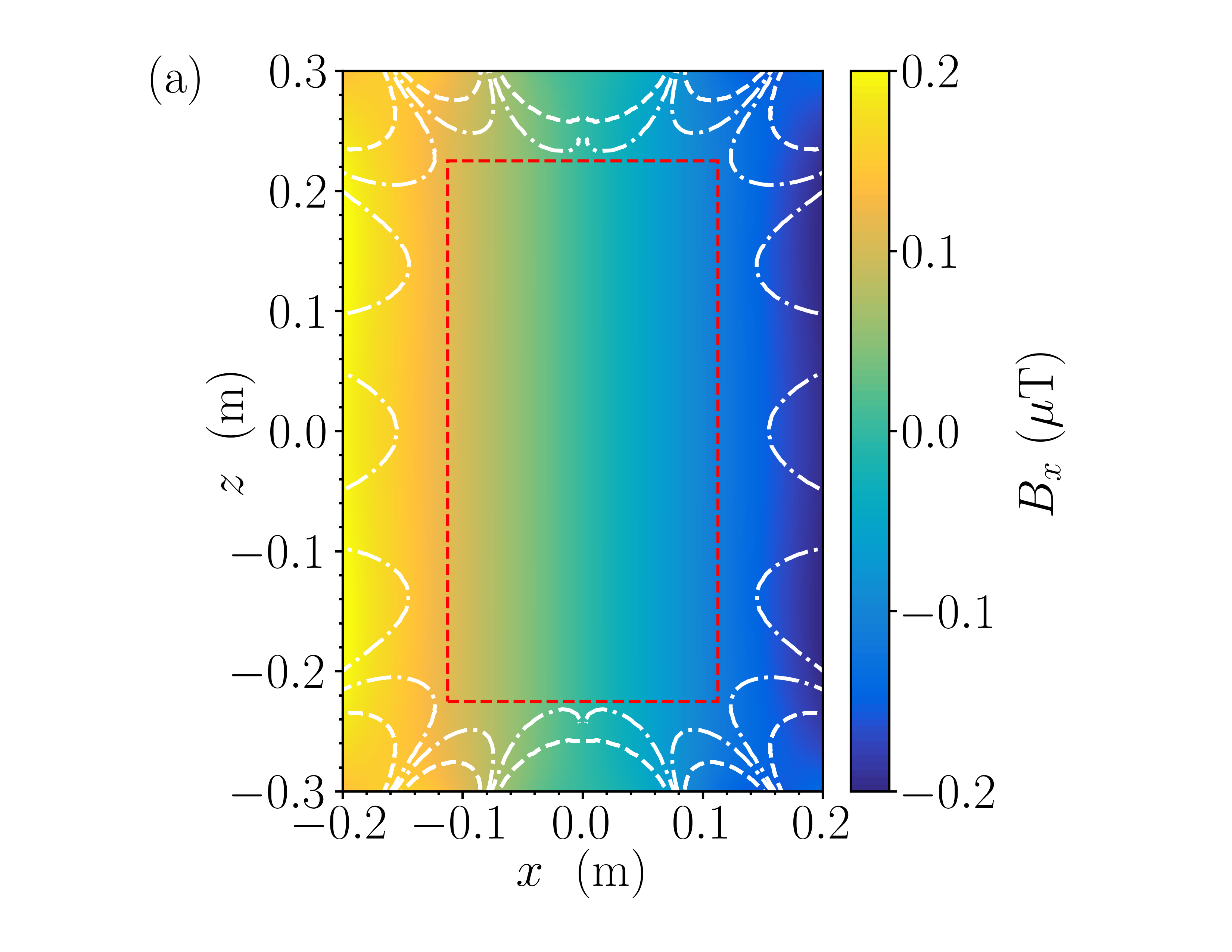}
         \caption{}
         \label{fig.bzcolormapsx}
     \end{subfigure}
     \begin{subfigure}{0.49\textwidth}
         \centering
         \includegraphics[width=\textwidth]{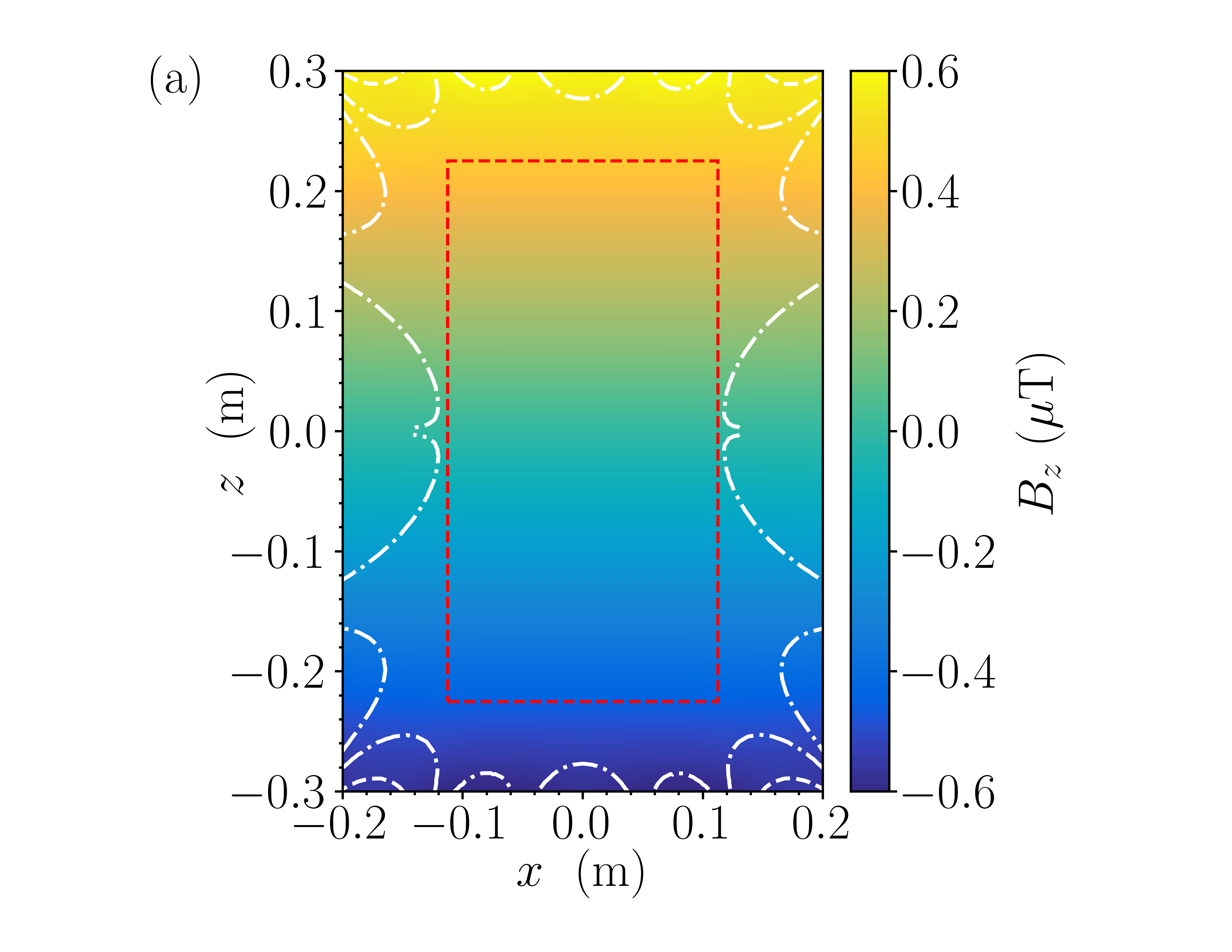}
         \caption{}
         \label{fig.bzcolormapsz}
     \end{subfigure}
\end{center}
\caption{Color maps showing the magnitude of the magnetic field, in the $xz$-plane inside a closed finite length perfect magnetic conductor generated by the active--passive system depicted in Fig.~\ref{fig.bzcoil}: (a) transverse component, $B_x$, and (b) axial component, $B_z$. The field profiles were calculated numerically using COMSOL Multiphysics\textsuperscript{\textregistered} Version 5.3a. The magnetic field is optimized between $\rho=[0,\rho_c/4]$ and $z=\pm{z'/2}$; dashed red lines in (a--b). Contours show where the field deviates from the target field by $5$\% (dashed curves) and $1$\% (dot-dashed curves).}
\label{fig.bzcolormaps}
\end{figure}
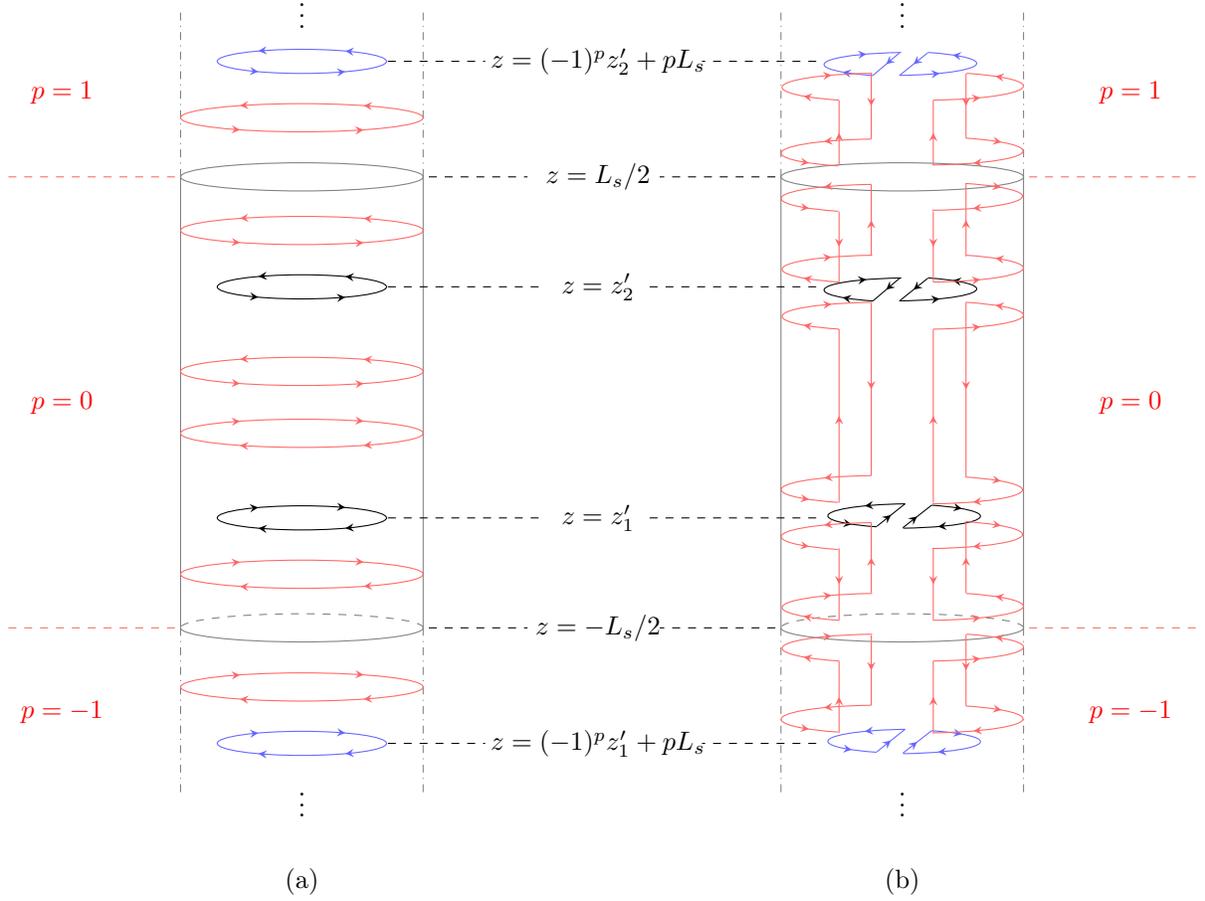
\begin{figure}[ht!]
\centering
\captionsetup[subfigure]{labelformat=empty}
\hspace{0pt}
\begin{subfigure}{0.49\textwidth}
\centering
\begin{tikzpicture}[scale=0.75]
\draw [gray,fill=gray!0] (0,4) ellipse (2.15 and 0.25);
\draw [gray] (-2.15,-4) arc (180:360:2.15 and 0.25);
\draw [dashed,gray] (2.15,-4) arc (0:180:2.15 and 0.25);
\draw [gray] (-2.15,-4) -- (-2.15,4);
\draw [gray] (2.15,-4) -- (2.15,4);
\draw [dash dot,gray] (-2.15,4) -- (-2.15,7);
\draw [dash dot,gray] (2.15,4) -- (2.15,7);
\draw [dash dot,gray] (-2.15,-4) -- (-2.15,-7);
\draw [dash dot,gray] (2.15,-4) -- (2.15,-7);
\node at (0,7) {$\vdots$};
\node at (0,-7) {$\vdots$};
\node at (0,-8.5) {$(\textnormal{a})$};

\draw [black,postaction={on each segment={mid arrow=black}}] (1.5,1.5*0.866+0.75) arc (0:360:1.5 and 0.214);
\draw [black,postaction={on each segment={mid arrow=black}}] (1.5,-1.5*0.866-0.75) arc (360:0:1.5 and 0.214);

\draw [blue!60,postaction={on each segment={mid arrow=blue!60}}] (1.5,1.5*0.866+4.75) arc (0:360:1.5 and 0.214);
\draw [blue!60,postaction={on each segment={mid arrow=blue!60}}] (1.5,-1.5*0.866-4.75) arc (360:0:1.5 and 0.214);

\draw [red!60,postaction={on each segment={mid arrow=red!60}}] (2.15,1.5*0.866+1.75) arc (0:360:2.15 and 0.25);
\draw [red!60,postaction={on each segment={mid arrow=red!60}}] (2.15,-1.5*0.866-1.75) arc (360:0:2.15 and 0.25);

\draw [red!60,postaction={on each segment={mid arrow=red!60}}] (2.15,1.5*0.866+3.75) arc (0:360:2.15 and 0.25);
\draw [red!60,postaction={on each segment={mid arrow=red!60}}] (2.15,-1.5*0.866-3.75) arc (360:0:2.15 and 0.25);

\draw [red!60,postaction={on each segment={mid arrow=red!60}}] (2.15,1.5*0.866-0.75) arc (0:360:2.15 and 0.25);
\draw [red!60,postaction={on each segment={mid arrow=red!60}}] (2.15,-1.5*0.866+0.75) arc (360:0:2.15 and 0.25);

\node at (5.25,-4) {$z=-L_s/2$};
\node at (5.25,4) {$z=L_s/2$};
\node at (-4.6,0) {$\textcolor{white}{stuff}$};
\draw [dashed] (2.25,4) -- (4,4);
\draw [dashed] (2.25,-4) -- (4,-4);

\node at (5.25,-6.05) {$z=(-1)^pz_1'+pL_s$};
\node at (5.25,6.05) {$z=(-1)^pz_2'+pL_s$};
\draw [dashed] (1.525,6.05) -- (3.25,6.05);
\draw [dashed] (1.525,-6.05) -- (3.25,-6.05);

\node at (5.25,-2.05) {$z=z_1'$};
\node at (5.25,2.05) {$z=z_2'$};
\draw [dashed] (1.525,2.05) -- (4.25,2.05);
\draw [dashed] (1.525,-2.05) -- (4.25,-2.05);

\draw [dashed,red!60] (-2.25,4) -- (-5.25,4);
\draw [dashed,red!60] (-2.25,-4) -- (-5.25,-4);

\node [red] at (-4.25,0) {$p=0$};
\node [red] at (-4.25,-5.5) {$p=-1$};
\node [red] at (-4.25,5.5) {$p=1$};
\end{tikzpicture}
\caption{}
\end{subfigure}
\hspace{-40pt}
\begin{subfigure}{0.49\textwidth}
\centering
\begin{tikzpicture}[scale=0.75]
\draw [gray,fill=gray!0] (0,4) ellipse (2.15 and 0.25);
\draw [gray] (-2.15,-4) arc (180:360:2.15 and 0.25);
\draw [dashed,gray] (2.15,-4) arc (0:180:2.15 and 0.25);
\draw [gray] (-2.15,-4) -- (-2.15,4);
\draw [gray] (2.15,-4) -- (2.15,4);
\draw [dash dot,gray] (-2.15,4) -- (-2.15,7);
\draw [dash dot,gray] (2.15,4) -- (2.15,7);
\draw [dash dot,gray] (-2.15,-4) -- (-2.15,-7);
\draw [dash dot,gray] (2.15,-4) -- (2.15,-7);
\node at (0,7) {$\vdots$};
\node at (0,-7) {$\vdots$};
\node at (0,-8.5) {$(\textnormal{b})$};

\draw [black,postaction={on each segment={mid arrow=black}}] (-0.52,1.5*0.866+0.5) arc (245:95:1.5 and 0.214);
\draw [black,postaction={on each segment={mid arrow=black}}] (0.52,-1.325-0.5) arc (425:275:1.5 and 0.214);
\draw [black,postaction={on each segment={mid arrow=black}}] (0.05,-1.8) arc (95:245:1.5 and 0.214);
\draw [black,postaction={on each segment={mid arrow=black}}]    (-0.05,1.8) arc (275:425:1.5 and 0.214);
\draw [black,postaction={on each segment={mid arrow=black}}] (-0.05,1.5*0.866+0.9) -- (-0.52,1.5*0.866+0.5);
\draw [black,postaction={on each segment={mid arrow=black}}] (0.45,1.5*0.866+0.9) -- (-0.05,1.5*0.866+0.5);
\draw [black,postaction={on each segment={mid arrow=black}}] 
(-0.45,-1.5*0.866-0.9) -- (0.05,-1.5*0.866-0.5);
\draw [black,postaction={on each segment={mid arrow=black}}] (0.01,-1.5*0.866-0.925) -- (0.52,-1.325-0.5);

\draw [blue!60,postaction={on each segment={mid arrow=blue!60}}] (-0.52,1.5*0.866+4.5) arc (245:95:1.5 and 0.214);
\draw [blue!60,postaction={on each segment={mid arrow=blue!60}}] (0.52,-1.325-4.5) arc (425:275:1.5 and 0.214);
\draw [blue!60,postaction={on each segment={mid arrow=blue!60}}] (0.05,-5.8) arc (95:245:1.5 and 0.214);
\draw [blue!60,postaction={on each segment={mid arrow=blue!60}}]    (-0.05,5.8) arc (275:425:1.5 and 0.214);
\draw [blue!60,postaction={on each segment={mid arrow=blue!60}}] (-0.05,1.5*0.866+4.9) -- (-0.52,1.5*0.866+4.5);
\draw [blue!60,postaction={on each segment={mid arrow=blue!60}}] (0.45,1.5*0.866+4.9) -- (-0.05,1.5*0.866+4.5);
\draw [blue!60,postaction={on each segment={mid arrow=blue!60}}] 
(-0.45,-1.5*0.866-4.9) -- (0.05,-1.5*0.866-4.5);
\draw [blue!60,postaction={on each segment={mid arrow=blue!60}}] (0.01,-1.5*0.866-4.925) -- (0.52,-1.325-4.5);

\draw [red!60,postaction={on each segment={mid arrow=red!60}}] (-1.42+0.3,1.5*0.866) arc (245:95:1.75 and 0.25);
\draw [red!60,postaction={on each segment={mid arrow=red!60}}] (1.420-0.3,-1.325) arc (425:275:1.75 and 0.25);
\draw [red!60,postaction={on each segment={mid arrow=red!60}}] (-2*0.45+0.35,-1.5*0.866) arc (95:245:1.75 and 0.25);
\draw [red!60,postaction={on each segment={mid arrow=red!60}}]    (0.9-0.35,1.5*0.866) arc (275:425:1.75 and 0.25);

\draw [red!60,postaction={on each segment={mid arrow=red!60}}] (-2*0.45+0.35,1.76) --  (-2*0.45+0.35,-1.5*0.866);
\draw [red!60,postaction={on each segment={mid arrow=red!60}}] (2*0.45-0.35,-2*0.89) --  (2*0.45-0.35,1.5*0.866);
\draw [red!60,postaction={on each segment={mid arrow=red!60}}] (2*0.71-0.29,1.78) -- (2*0.71-0.29,-1.5*0.88);
\draw [red!60,postaction={on each segment={mid arrow=red!60}}] (-2*0.71+0.3,-2*0.89) -- (-2*0.71+0.3,1.5*0.866);

\draw [red!60,postaction={on each segment={mid arrow=red!60}}] (-1.42+0.3,-1.5*0.866/1.5+3) arc (245:95:1.75 and 0.25);
\draw [red!60,postaction={on each segment={mid arrow=red!60}}] (1.420-0.3,1.325/1.5+3) arc (425:275:1.75 and 0.25);
\draw [red!60,postaction={on each segment={mid arrow=red!60}}] (-2*0.45+0.35,1.5*0.866/1.5+3) arc (95:245:1.75 and 0.25);
\draw [red!60,postaction={on each segment={mid arrow=red!60}}]    (0.9-0.35,-1.5*0.866/1.5+3) arc (275:425:1.75 and 0.25);

\draw [red!60,postaction={on each segment={mid arrow=red!60}}] (-2*0.45+0.35,-1.76/1.5+3.775) --  (-2*0.45+0.35,1.5*0.866/1.5+3);
\draw [red!60,postaction={on each segment={mid arrow=red!60}}] (2*0.45-0.35,2*0.89/1.5+2.2) --  (2*0.45-0.35,-1.5*0.866/1.5+3);
\draw [red!60,postaction={on each segment={mid arrow=red!60}}] (2*0.71-0.29,-1.78/1.5+3.775) -- (2*0.71-0.29,1.5*0.88/1.5+3);
\draw [red!60,postaction={on each segment={mid arrow=red!60}}] (-2*0.71+0.3,2*0.89/1.5+2.2) -- (-2*0.71+0.3,-1.5*0.866/1.5+3);

\draw [red!60,postaction={on each segment={mid arrow=red!60}}] (-1.42+0.3,-1.5*0.866/1.5-3) arc (245:95:1.75 and 0.25);
\draw [red!60,postaction={on each segment={mid arrow=red!60}}] (1.420-0.3,1.325/1.5-3) arc (425:275:1.75 and 0.25);
\draw [red!60,postaction={on each segment={mid arrow=red!60}}] (-2*0.45+0.35,1.5*0.866/1.5-3) arc (95:245:1.75 and 0.25);
\draw [red!60,postaction={on each segment={mid arrow=red!60}}]    (0.9-0.35,-1.5*0.866/1.5-3) arc (275:425:1.75 and 0.25);

\draw [red!60,postaction={on each segment={mid arrow=red!60}}] (-2*0.45+0.35,-1.76/1.5-2.2) --  (-2*0.45+0.35,1.5*0.866/1.5-3);
\draw [red!60,postaction={on each segment={mid arrow=red!60}}] (2*0.45-0.35,2*0.89/1.5-3.775) --  (2*0.45-0.35,-1.5*0.866/1.5-3);
\draw [red!60,postaction={on each segment={mid arrow=red!60}}] (2*0.71-0.29,-1.78/1.5-2.2) -- (2*0.71-0.29,1.5*0.88/1.5-3);
\draw [red!60,postaction={on each segment={mid arrow=red!60}}] (-2*0.71+0.3,2*0.89/1.5-3.775) -- (-2*0.71+0.3,-1.5*0.866/1.5-3);

\draw [red!60,postaction={on each segment={mid arrow=red!60}}] (-2*0.45+0.35,-1.76/1.5-4.2) arc (95:245:1.75 and 0.25);
\draw [red!60,postaction={on each segment={mid arrow=red!60}}] (2*0.45-0.35,2*0.89/1.5-5.775) arc (275:425:1.75 and 0.25);
\draw [red!60,postaction={on each segment={mid arrow=red!60}}] (-2*0.71+0.3,2*0.89/1.5-5.775) arc (245:95:1.75 and 0.25);
\draw [red!60,postaction={on each segment={mid arrow=red!60}}]    (2*0.71-0.29,-1.78/1.5-4.2) arc (425:275:1.75 and 0.25);

\draw [red!60,postaction={on each segment={mid arrow=red!60}}]  (-2*0.45+0.35,1.5*0.866/1.5-5) -- (-2*0.45+0.35,-1.76/1.5-4.2);
\draw [red!60,postaction={on each segment={mid arrow=red!60}}] (2*0.45-0.35,-1.5*0.866/1.5-5) -- (2*0.45-0.35,2*0.89/1.5-5.775);
\draw [red!60,postaction={on each segment={mid arrow=red!60}}] (2*0.71-0.29,1.5*0.88/1.5-5) -- (2*0.71-0.29,-1.78/1.5-4.2);
\draw [red!60,postaction={on each segment={mid arrow=red!60}}] (-2*0.71+0.3,-1.5*0.866/1.5-5) -- (-2*0.71+0.3,2*0.89/1.5-5.775);

\draw [red!60,postaction={on each segment={mid arrow=red!60}}] (-2*0.45+0.35,4.67) arc (95:245:1.75 and 0.25);
\draw [red!60,postaction={on each segment={mid arrow=red!60}}] (2*0.45-0.35,2*0.89/1.5+4.175) arc (275:425:1.75 and 0.25);
\draw [red!60,postaction={on each segment={mid arrow=red!60}}] (-2*0.71+0.3,2*0.89/1.5+4.175) arc (245:95:1.75 and 0.25);
\draw [red!60,postaction={on each segment={mid arrow=red!60}}]    (2*0.71-0.29,+1.78/1.5+3.5) arc (425:275:1.75 and 0.25);

\draw [red!60,postaction={on each segment={mid arrow=red!60}}]  (-2*0.45+0.35,1.5*0.866/1.5+4.975) -- (-2*0.45+0.35,1.76/1.5+3.5);
\draw [red!60,postaction={on each segment={mid arrow=red!60}}] (2*0.45-0.35,1.5*0.866/1.5+3.34) -- (2*0.45-0.35,2*0.89/1.5+4.17);
\draw [red!60,postaction={on each segment={mid arrow=red!60}}] (2*0.71-0.29,1.5*0.88/1.5+4.95) -- (2*0.71-0.29,1.78/1.5+3.5);
\draw [red!60,postaction={on each segment={mid arrow=red!60}}] (-2*0.71+0.3,1.5*0.866/1.5+3.34) -- (-2*0.71+0.3,2*0.89/1.5+4.17);

\node at (-4.5,0) {$\textcolor{white}{stuff}$};

\draw [dashed] (-1.525,2.05) -- (-4.55,2.05);
\draw [dashed] (-1.525,-2.05) -- (-4.55,-2.05);

\draw [dashed] (-1.525,6.05) -- (-3.55,6.05);
\draw [dashed] (-1.525,-6.05) -- (-3.55,-6.05);

\draw [dashed] (-2.25,4) -- (-4.3,4);
\draw [dashed] (-2.25,-4) -- (-4.3,-4);

\draw [dashed,red!60] (2.25,4) -- (5.25,4);
\draw [dashed,red!60] (2.25,-4) -- (5.25,-4);

\node [red] at (4.05,0) {$p=0$};
\node [red] at (4.05,-5.5) {$p=-1$};
\node [red] at (4.05,5.5) {$p=1$};
\end{tikzpicture}
\caption{}
\end{subfigure}
\caption{Schematic diagram showing how the approximate zonal (a) and tesseral (b) harmonic responses are generated by simple current loops with cylindrical and $m$-fold azimuthal symmetries, respectively. The discrete current loops (black) are located at $z=z_1'$ and $z=z_2'$ with their $p^\textnormal{th}$ reflected pseudo-image-currents (red) generated by the surface of the cylindrical shield at $\rho=\rho_s$ and by the planar end caps at $z=\pm L_s/2$, in accordance with the method of images described by \eqref{eq.pseudo}. Images of the two planar (black) coils resulting from the end caps are located at $z=(-1)^pz_1'+pL_s$ and $z=(-1)^pz_2'+pL_s$, respectively, where $p\in \mathbb{Z}$ (two such image coils are shown blue). For all odd reflections the axial current direction is reversed.}
\label{fig.responsediagram}
\end{figure}
\section{Conclusion}
Here we have formulated an analytical model to calculate the magnetic field generated by an arbitrary static current distribution confined to a plane whose normal is parallel to the axis of a finite closed high-permeability cylinder. Our formalism is based on a Green's function expansion of the vector potential generated by a planar coil system. To satisfy the boundary conditions at the surface of the cylinder, we assume that it is a perfect magnetic conductor and express the induced magnetization in terms of a pseudo-current density. Due to the shared azimuthal symmetries of the planar current flows and the induced pseudo-current on the cylindrical surface of the magnetic shield, the response of the magnetic shield can be expressed in terms of the planar current distribution. We used this formalism to enable a priori optimization of the planar current distribution required to generate specified target magnetic field profiles by combining a Fourier--Bessel decomposition of the current flow with a quadratic minimization procedure.

We demonstrated this optimization methodology by using it to design bi-planar coils that accurately generate a constant transverse field, $B_x$, across the cylinder, and a linear gradient field, $\mathrm{d}B_z/\mathrm{d}z$, along the axis of the cylinder. Predictions from our analytical model agree well with with subsequent forward finite element simulations, validating our methodology. We quantified the interaction of planar systems with a closed finite high-permeability cylinder and found that all coil designs without complete azimuthal symmetry induced magnetic fields in the cylindrical surface of the shield that opposed the field generated by the planar current source. Conversely, planar end caps have the opposite effect and amplify the field from the planar current source.

Our analytical model and optimization procedure extends the range of current geometries that can be tailored within finite closed cylindrical magnetic shields in order to accurately generate specified target field profiles. This enables the development of systems that require the best magnetic field control that can be achieved subject to size, weight, power and cost constraints. It may also enable hybrid active/passive planar coils to be retrofitted to existing cylindrical magnetic shields in order to improve their performance for a low cost and without disrupting existing experimental setups. In the future, combining planar and cylindrical coils could enable ultra-precise magnetic fields to be generated through even larger interior fractions of magnetic shields.
\\
The PYTHON code used to design arbitrary planar coils in a magnetically shielded cylinder is openly available from GitHub \cite{pjpython}. Verification using COMSOL Multiphysics requires a valid license.

\section*{Acknowledgments} 
We acknowledge support from the UK Quantum Technology Hub Sensors and Timing, funded by the Engineering and Physical Sciences Research Council (EP/M013294/1). \\
The authors M.P., P.J.H., T.M.F., M.J.B., and R.B. declare that they have a patent pending to the UK Government Intellectual Property Office (Application No.1913549.0) regarding the magnetic field optimization techniques described in this work. N.H, M.J.B, and R.B also declare financial interests in a University of Nottingham spin-out company, Cerca. The authors have no other competing financial interests.

\section*{Appendix A: Magnetic Field Derivation}
 \label{app.mag}
In order to calculate the magnetic field, we must first determine both the Fourier transform of the streamfunction and the induced Fourier pseudo-current density. Substituting \eqref{eq.streamy} into \eqref{eq.ht} and integrating over the azimuthal angle yields 
\begin{align}\nonumber
    \varphi^m(k)=&\frac{\rho_c}{2}\sum_{n=1}^N\sum_{m'=0}^M\left(W_{nm'}\left(\delta_{mm'}+\delta_{m,-m'}\right)+iQ_{nm'}\left(\delta_{mm'}-\delta_{m,-m'}\right)\right)\\ & \hspace{230pt} \times
    \int_0^{\rho_c}\mathrm{d}\rho' \ \rho'J_m(k\rho')J_{m'}\left(\frac{\rho_{nm'}\rho'}{\rho_c}\right).\label{eq.streamy2}
\end{align}
Upon substituting \eqref{eq.streamy2} into the expressions for the magnetic field, \eqref{eq.shieldyBr}-\eqref{eq.shieldyBz}, the $\delta-$functions collapse the infinite summation, resulting in every Bessel function of order $m$ being replaced by one of order $m'$. Hence, without loss of generality, applying a Bessel function product identity (5.54.1 in \cite{gradshteyn2007}), we may now write
\begin{align}\nonumber
    \varphi^m(k)=&\frac{\rho_c^3}{2}\sum_{n=1}^N\sum_{m'=0}^M\left(W_{nm'}\left(\delta_{mm'}+\delta_{m,-m'}\right)+iQ_{nm'}\left(\delta_{mm'}-\delta_{m,-m'}\right)\right)\\ & \hspace{230pt} \times
 \frac{\rho_{nm'}}{k^2\rho_c^2-\rho_{nm'}^2}J_{m'}(k\rho_c)J'_{m'}\left(\rho_{nm'}\right).\label{eq.streamy3}
\end{align}
Similarly, we may also calculate the induced Fourier pseudo-current density. Substituting \eqref{eq.streamy} into \eqref{eq.pseudo} and integrating over the azimuthal coordinate yields
\begin{align}\nonumber
    J_\phi^{mp}(k)=\frac{\rho_ce^{-ik\left((-1)^pz'+pL_s\right)}}{2\rho_s|k|I_m\left(|k|\rho_s\right)K'_m\left(|k|\rho_s\right)}\sum_{n=1}^N\sum_{m'=0}^M\left(W_{nm'}\left(\delta_{mm'}+\delta_{m,-m'}\right)+iQ_{nm'}\left(\delta_{mm'}-\delta_{m,-m'}\right)\right) \\ \times \int_0^\infty \mathrm{d}\tilde{k} \ \frac{\tilde{k}^3}{\tilde{k}^2+k^2} J_m(\tilde{k}\rho_s) \int_0^{\rho_c}\mathrm{d}\rho' \ \rho'J_{m'}(\tilde{k}\rho')J_{m'}\left(\frac{\rho_{nm'}\rho'}{\rho_c}\right).\label{eq.pseudo1}
\end{align}
To simplify these expressions, we evaluate the integral over $\tilde{k}$ first and then separate the integrand using partial fractions to yield
\begin{align} \label{eq.Jsep}
    \int_0^\infty \mathrm{d}\tilde{k} \ \frac{\tilde{k}^3}{\tilde{k}^2+k^2} J_m(\tilde{k}\rho_s)J_m(\tilde{k}\rho')=\int_0^\infty \mathrm{d}\tilde{k} \ \tilde{k}J_m(\tilde{k}\rho_s)J_m(\tilde{k}\rho')-k^2\int_0^\infty \mathrm{d}\tilde{k} \ \frac{\tilde{k}}{\tilde{k}^2+k^2} J_m(\tilde{k}\rho_s)J_m(\tilde{k}\rho').
\end{align}
Inspecting \eqref{eq.Jsep} we see that the first component is simply the orthogonality relation between Bessel functions, and the second has a known solution (6.541.1 in \cite{gradshteyn2007}). Hence, we can state that
\begin{align}\label{eq.ind}
\int_0^\infty \mathrm{d}\tilde{k} \ \frac{\tilde{k}^3}{\tilde{k}^2+k^2} J_m(\tilde{k}\rho_s)J_m(\tilde{k}\rho')
= \frac{\delta{\left(\rho'-\rho_s\right)}}{\rho'} - k^2I_{m'}\left(|k|\rho'\right)K_{m'}\left(|k|\rho_s\right).
\end{align}
However, since $\rho'<\rho_s$, the first term in \eqref{eq.ind} can be neglected. Inserting \eqref{eq.ind} into \eqref{eq.pseudo1} and integrating over $\rho'$ results in the expression
\begin{align}\nonumber
    J_\phi^{mp}(k)=\frac{\rho_ck^2e^{-ik\left((-1)^pz'+pL_s\right)}}{2\rho_s|k|I_m\left(|k|\rho_s\right)K'_m\left(|k|\rho_s\right)}\sum_{n=1}^N\sum_{m'=0}^M\left(W_{nm'}\left(\delta_{mm'}+\delta_{m,-m'}\right)+iQ_{nm'}\left(\delta_{mm'}-\delta_{m,-m'}\right)\right) \\ \times  \frac{\rho_{nm'}}{k^2\rho_c^2+\rho_{nm'}^2}J'_{m'}\left(\rho_{nm'}\right)I_{m'}\left(|k|\rho_c\right)K_{m'}\left(|k|\rho_s\right).\label{eq.pseudo2}
\end{align}
Substituting \eqref{eq.streamy3} and \eqref{eq.pseudo2} into \eqref{eq.shieldyBr}-\eqref{eq.shieldyBz}, we can express the summation over the exponentials associated with the infinite reflections of the planar streamfuction as
\begin{align} \label{eq.summyboy1}
    \sum_{p=-\infty}^\infty e^{-k|z-(-1)^pz'+pL_s|} 
    &= e^{-k|z-z'|} + \frac{2}{e^{2kL}-1}\big(e^{kL}\cosh\left(k\left(z+z'\right)\right) + \cosh\left(k\left(z-z'\right)\right)\big),
\end{align}
and
\begin{align}
    \sum_{p=-\infty}^\infty \frac{\left(z-(-1)^pz'+pL_s\right)}{\left|z-(-1)^pz'+pL_s\right|}e^{-k|z-(-1)^pz'+pL_s|}  &=
    \frac{(z-z')}{|z-z'|} e^{-k|z-z'|} \nonumber \\ & \hspace{10pt}- \frac{2}{e^{2kL}-1}\big(e^{kL}\sinh\left(k\left(z+z'\right)\right)+\sinh\left(k\left(z-z'\right)\right)\big).
\end{align}
Similarly, the summation over the infinite pseudo-current reflections can be written as a Fourier series expansion,  
\begin{align} \label{eq.diracyboy}
    \sum_{p=-\infty}^\infty  e^{ik\left(z-(-1)^pz'+pL_s\right)} &= \frac{\pi}{L} \sum_{p=-\infty}^\infty \delta\left(k-\frac{p\pi}{L}\right) \left(e^{ik(z+z'-L)} +e^{ik(z-z')} \right).
\end{align}
Using these expansions we may finally write
\begin{equation}
    B_{\rho}\left(\rho,\phi,z\right)=\frac{\mu_0\rho_c^3}{2}  \sum_{n=1}^{N}\sum_{m=0}^{M}\rho_{nm}J'_m(\rho_{nm})\left(W_{nm}\cos\left(m\phi\right)+Q_{nm}\sin\left(m\phi\right)\right)B^{nm}_\rho(\rho,z),
\end{equation}
\begin{equation}
    B_{\phi}\left(\rho,\phi,z\right)=\frac{\mu_0\rho_c^3}{2\rho} \sum_{n=1}^{N}\sum_{m=0}^{M}m\rho_{nm}J'_m(\rho_{nm})\left(W_{nm}\sin\left(m\phi\right)-Q_{nm}\cos\left(m\phi\right)\right)B^{nm}_\phi(\rho,z),
\end{equation}
\begin{equation}
    B_{z}\left(\rho,\phi,z\right)=\frac{\mu_0\rho_c^3}{2} \sum_{n=1}^N\sum_{m=0}^{M}\rho_{nm}J'_m(\rho_{nm})\left(W_{nm}\cos\left(m\phi\right)+Q_{nm}\sin\left(m\phi\right)\right)B^{nm}_z(\rho,z),
\end{equation}
where
\begin{align}
    B^{nm}_\rho(\rho,z)=
    &-\int_0^\infty \mathrm{dk} \ k^2 \sigma\left(k;z,z',L_s\right) \frac{J'_m(k\rho)J_m(k\rho_c)}{k^2\rho_c^2-\rho_{nm}^2}\nonumber \\[6pt]
    &\hspace{140pt} - \sum_{p=1}^\infty \tilde{p}\left|\tilde{p}\right|
    \lambda_p\left(z,z',L_s\right)
    \frac{I'_m(\left|\tilde{p}\right|\rho)I_m(\left|\tilde{p}\right|\rho_c)K_m(\left|\tilde{p}\right|\rho_s)}{I_m(\left|\tilde{p}\right|\rho_s)(\left|\tilde{p}\right|^2\rho_c^2+\rho_{nm}^2)},
\end{align}
\begin{align}
    B^{nm}_\phi(\rho,z)=
    &\int_0^\infty \mathrm{d}k \ k \sigma\left(k;z,z',L_s\right) \frac{J_m(k\rho)J_m(k\rho_c)}{k^2\rho_c^2-\rho_{nm}^2}\nonumber \\[6pt]
    &\hspace{140pt} + \sum_{p=1}^\infty \tilde{p} \ \lambda_p\left(z,z',L_s\right)
    \frac{I_m(\left|\tilde{p}\right|\rho)I_m(\left|\tilde{p}\right|\rho_c)K_m(\left|\tilde{p}\right|\rho_s)}{I_m(\left|\tilde{p}\right|\rho_s)(\left|\tilde{p}\right|^2\rho_c^2+\rho_{nm}^2)},
\end{align}
\begin{align}\nonumber
    B^{nm}_z(\rho,z)=
    &\int_0^\infty \mathrm{d}k \ k^2 \gamma\left(k;z,z',L_s\right) \frac{J_m(k\rho)J_m(k\rho_c)}{k^2\rho_c^2-\rho_{nm}^2} \\[6pt]
    &\hspace{140pt} - \sum_{p=1}^\infty \tilde{p}^2 \
    \tau_p\left(z,z',L_s\right)
    \frac{I_m(\left|\tilde{p}\right|\rho)I_m(\left|\tilde{p}\right|\rho_c)K_m(\left|\tilde{p}\right|\rho_s)}{I_m(\left|\tilde{p}\right|\rho_s)(\left|\tilde{p}\right|^2\rho_c^2+\rho_{nm}^2)},
\end{align}
where
\begin{equation}
    \gamma\left(k;z,z',L_s\right) = e^{-k|z-z'|} + \frac{2}{e^{2kL_s}-1}\bigg[e^{kL_s}\cosh\left(k\left(z+z'\right)\right) + \cosh\left(k\left(z-z'\right)\right)\bigg],
\end{equation}
\begin{equation}
    \sigma\left(k;z,z',L_s\right) = \frac{(z-z')}{|z-z'|} e^{-k|z-z'|} - \frac{2}{e^{2kL_s}-1}\bigg[e^{kL_s}\sinh\left(k\left(z+z'\right)\right) + \sinh\left(k\left(z-z'\right)\right)\bigg],
\end{equation}
\begin{align}
    \lambda_p\left(z,z',L_s\right) &= \frac{2}{L_s}\left(\left(-1\right)^p\sin\left(\tilde{p}(z+z')\right) + \sin\left(\tilde{p}(z-z')\right)\right), \\[8pt]
    \tau_p\left(z,z',L_s\right) &= \frac{2}{L_s}\left(\left(-1\right)^p\cos\left(\tilde{p}(z+z')\right) + \cos\left(\tilde{p}(z-z')\right)\right),
\end{align}
and $\tilde{p}=p\pi/L_s$.
\section*{Appendix B: Power Dissipation} \label{app.pwr}
The power dissipated in the surface of a circular planar current source of thickness $t$ and resistivity $\varrho$ is given by
\begin{equation} \label{eq.powernew1}
    P=\frac{\varrho}{t}\int_{0}^{\rho_c}\mathrm{d}\rho'\rho'\int_{0}^{2\pi}\mathrm{d}\phi'\ |J_{\rho}(\rho',\phi')|^2+|J_\phi(\rho',\phi')|^2.
\end{equation}
We can substitute the streamfunction \eqref{eq.streamy} into the continuity relations \eqref{eq.streamyd} to obtain
\begin{align}\label{eq.jrho}
    J_\rho\left(\rho',\phi'\right)=&\left(H\left(\rho'-\rho_c\right)-H\left(\rho'\right)\right)\frac{\rho_c}{\rho'} \sum_{n=1}^N\sum_{m=0}^MmJ_m\left(\frac{\rho_{nm}\rho'}{\rho_c}\right) \left(W_{nm}\sin(m\phi')-Q_{nm}\cos(m\phi')\right),
\end{align}
\begin{align}\label{eq.jphi}
    J_\phi\left(\rho',\phi'\right)=&\left(H\left(\rho'-\rho_c\right)-H\left(\rho'\right)\right) \sum_{n=1}^N\sum_{m=0}^M\rho_{nm}J'_m\left(\frac{\rho_{nm}\rho'}{\rho_c}\right) \left(W_{nm}\cos(m\phi')+Q_{nm}\sin(m\phi')\right).
\end{align}
Inserting \eqref{eq.jrho}-\eqref{eq.jphi} into \eqref{eq.powernew1} and integrating over the azimuthal component results in
\begin{align}\label{eq.powernewform1}
    P=&\frac{\varrho}{t}\pi\rho_c^2\sum_{n=1}^N\sum_{m=0}^M\left[\left(1+\delta_{m0}\right)W_{nm}^2+\left(1-\delta_{m0}\right)Q_{nm}^2\right] \nonumber \\ &\qquad\qquad\qquad\qquad\qquad\qquad \times \int_{0}^{1}\mathrm{d}\tilde{\rho} \
    \Bigg[ \frac{2m^2}{\tilde{\rho}}J_{m}\left(\rho_{nm}\tilde{\rho}\right)^2 - \rho_{nm}^2 \ \tilde{\rho} \ J_{m-1}\left(\rho_{nm}\tilde{\rho}\right) J_{m+1}\left(\rho_{nm}\tilde{\rho}\right) \Bigg],
\end{align}
which, when integrated over the radial component, can be expressed as
\begin{equation}
    P=\frac{\varrho}{t}\pi\rho_c^2\sum_{n=1}^N W_{n0}^2\rho_{n0}^2 J_1\left(\rho_{n0}\right)^2,
\end{equation} 
for $m=0$, and
\begin{align}\nonumber
    P=&\frac{\varrho}{t}\pi\rho_c^2\sum_{n=1}^N\sum_{m=1}^M\left(W_{nm}^2+Q_{nm}^2\right)
    \left(\frac{\rho_{nm}}{2}\right)^{2m}\frac{1}{m!(m-1)!} \\ 
    &\hspace{35pt} \Bigg[\ _2\tilde{F}_3\left(m,m+\frac{1}{2};m+1,m+1,2 m+1;-\rho_{nm}^2\right)\nonumber \\ 
    &\hspace{70pt} -\frac{\rho_{nm}^2}{2(m+1)^2} \ _3\tilde{F}_4\left(m+\frac{1}{2},m+1,m+1;m,m+2,m+2,2 m+1;-\rho_{nm}^2\right)\Bigg]\label{eq.powermmore1},
\end{align}
for $m \in \mathbb{Z}^+$, where $_i\tilde{F}_j$ is the regularized hypergeometric function. 
Some specific evaluations of \eqref{eq.powermmore1} are:
\begin{equation}
    m=1: \hspace{90pt} P=\frac{\varrho}{2t}\pi\rho_c^2\sum_{n=1}^N \left(W_{n1}^2+Q_{n1}^2\right)\rho_{n1}^2 J_0\left(\rho_{n1}\right)^2, \hspace{104pt}
\end{equation}
\begin{align}
    m=2: \qquad P=&\frac{\varrho}{2t}\pi\rho_c^2\sum_{n=1}^N \left(W_{n2}^2+Q_{n2}^2\right)
  \Bigg[\left(\rho_{n2}^2-4\right) J_0\left(\rho_{n2}\right)^2 -\frac{2 \left(\rho_{n2}^2-8\right)}{\rho_{n2}} J_0\left(\rho_{n2}\right) J_1\left(\rho_{n2}\right) \nonumber \\ 
  &\hspace{230pt} +\frac{\left(\rho_{n2}^4-16\right)}{\rho_{n2}^2} J_1\left(\rho_{n2}\right)^2\Bigg],
\end{align}
\begin{align}
    m=3: \qquad P=&\frac{\varrho}{2t}\pi\rho_c^2\sum_{n=1}^N \left(W_{n3}^2+Q_{n3}^2\right)
  \Bigg[\frac{\left(\rho_{n3}^4-96\right)}{\rho_{n3}^2} J_0\left(\rho_{n3}\right)^2 -\frac{48 \left(\rho_{n3}^2-8\right)}{\rho_{n3}^3} J_0\left(\rho_{n3}\right) J_1\left(\rho_{n3}\right) \nonumber \\ &\hspace{155pt} +\frac{\left(\rho_{n3}^6 -10\rho_{n3}^4 + 96\rho_{n3}^2 -384\right)}{\rho_{n3}^4} J_1\left(\rho_{n3}\right)^2 \Bigg].
\end{align}
\section*{Appendix C: Example Coil Designs}
\begin{figure}[htb!]
\captionsetup[subfigure]{labelformat=empty}
\begin{center}
     \begin{subfigure}{0.49\textwidth}
         \centering
        \includegraphics[width=\textwidth]{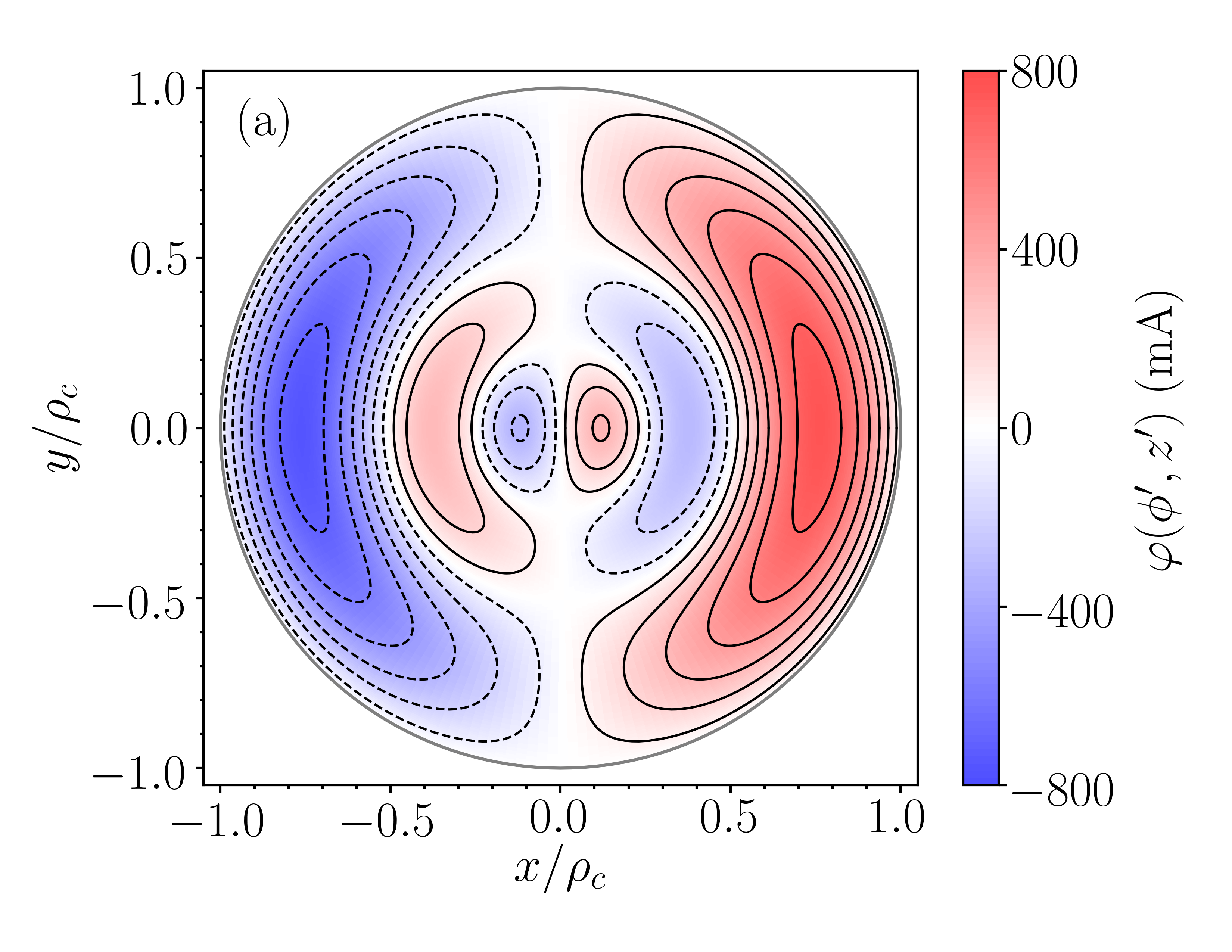}
         \caption{}
         \label{fig.dbxtop}
     \end{subfigure}
     \begin{subfigure}{0.49\textwidth}
         \centering
        \includegraphics[width=\textwidth]{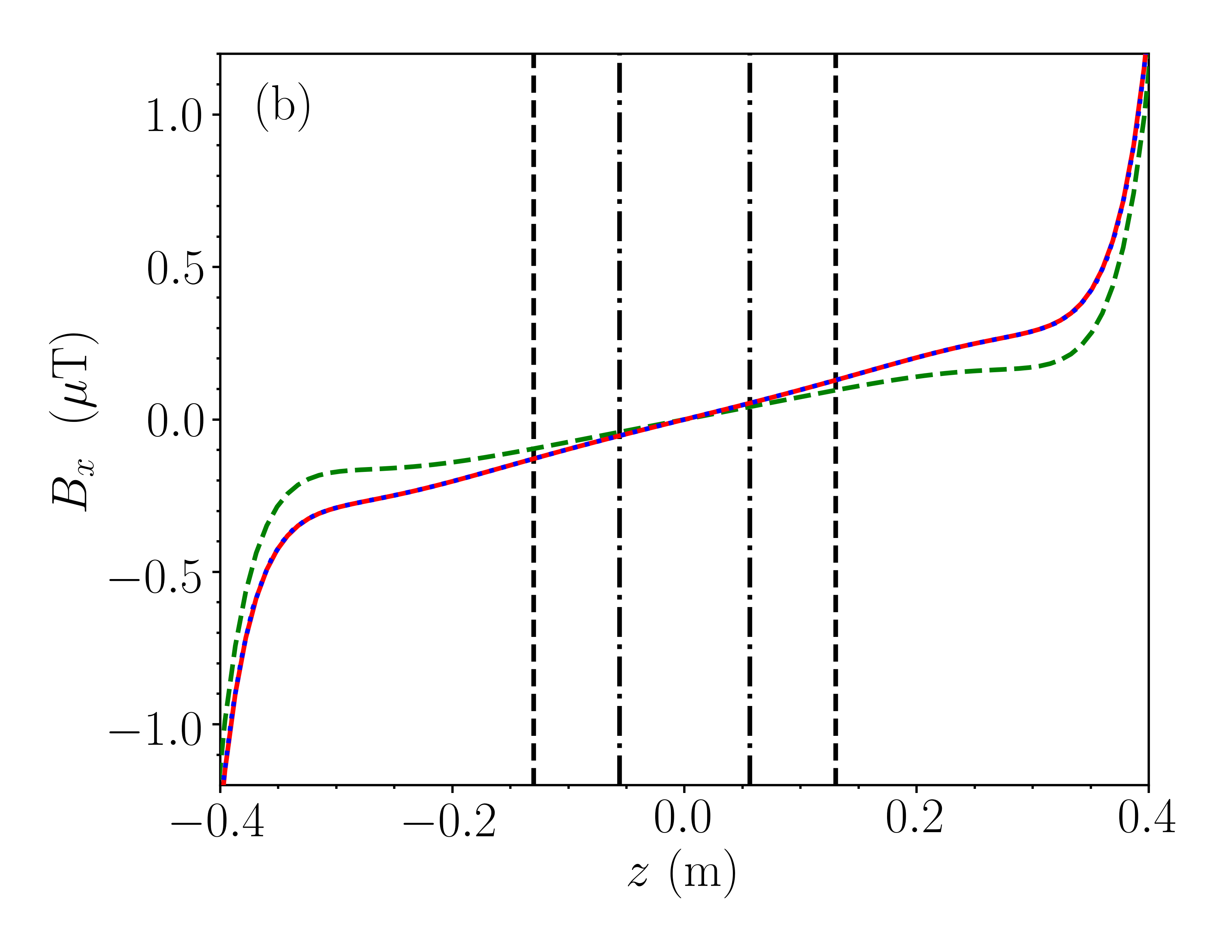}
         \caption{}
         \label{fig.dbxbottom}
     \end{subfigure}
\end{center}
\caption{Wire layouts (a) and performance (b) of a hybrid active--passive system optimized to generate to generate a linear variation of $B_x$ with $z$-position. Current flows on the surface of two disks of radius $\rho_c=0.45$ m, which are separated symmetrically from the origin and lie at $z'=\pm0.45$ m. The wire layouts are optimized to generate a linear transverse field gradient, $\mathrm{d}B_x/\mathrm{d}z=1$~$\mu$T/m, along the $z$-axis of the cylinder through the harmonic field profile $\mathbf{B}=(z~\mathbf{\hat{x}}+x~\mathbf{\hat{z}})$. The current-carrying planes are placed symmetrically inside a perfect closed magnetic shield of radius $\rho_s=0.5$ m and length $L_s=1$ m and the magnetic field is optimized along the $z$-axis between $z=\pm{z'/2}$. The least squares optimization was performed with parameters $N=100$, $M=1$, $\beta=1.77\times10^{-8}$~T\textsuperscript{2}/W, $t=0.5$~mm, and $\rho=1.68\times10^{-8}$~$\Omega$m. (a) Color map of the optimal current streamfunction calculated for the upper current-carrying plane in Fig.~\ref{fig.shieldandcoil} [blue and red shaded regions correspond to current counter flows and their intensity shows the streamfunction magnitude from low (white) to high (intense color)]. Solid and dashed black curves represent discrete wires with opposite senses of current flow, approximating the current continuum with $N_\varphi=12$ global contour levels. Streamfunction on the lower light brown plane in Fig.~\ref{fig.shieldandcoil} is geometrically identical but the current direction is reversed. (b) Transverse magnetic field, $B_x$, calculated versus axial position, $z$, from the current continuum in (a) in three ways: analytically using \eqref{eq.cbrff}-\eqref{eq.cbzff} (solid red curve); numerically using COMSOL Multiphysics\textsuperscript{\textregistered} Version 5.5a, modeling the high-permeability cylinder as a perfect magnetic conductor (blue dotted curve); numerically \emph{without} the high-permeability cylinder and using the Biot–Savart law with $N_\varphi=100$ contour levels (dashed green curve). Black lines enclose the regions where the axial field gradient deviates by $5$\% (dashed) and $1$\% (dot-dashed).}
\label{fig.dbxcoil}
\end{figure}

\begin{figure}[htb!]
\captionsetup[subfigure]{labelformat=empty}
\begin{center}
     \begin{subfigure}{0.49\textwidth}
         \centering
        \includegraphics[width=\textwidth]{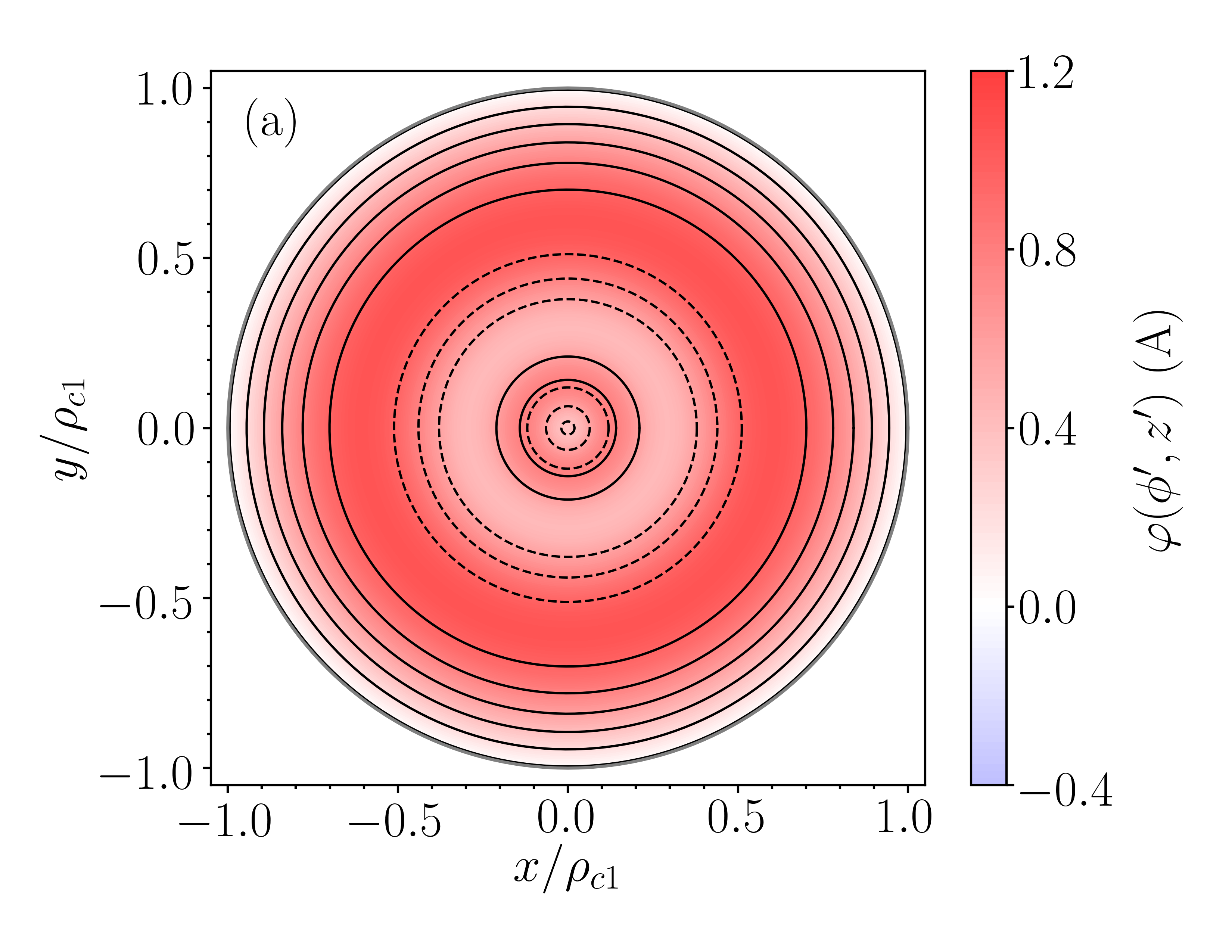}
         \caption{}
         \label{fig.bztop}
     \end{subfigure}
    \begin{subfigure}{0.49\textwidth}
         \centering
        \includegraphics[width=\textwidth]{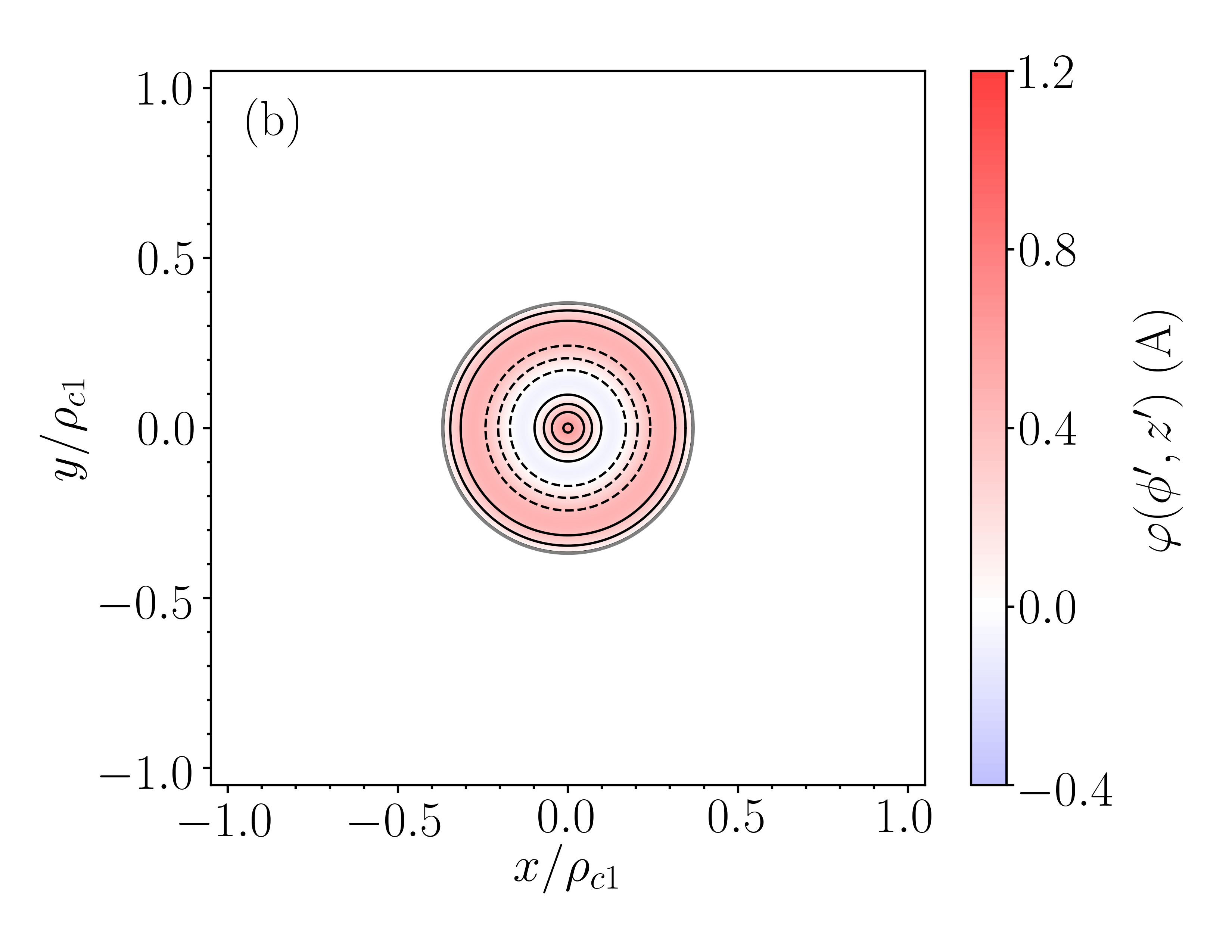}
         \caption{}
         \label{fig.bzbottom}
     \end{subfigure}
     \begin{subfigure}{0.49\textwidth}
         \centering
        \includegraphics[width=\textwidth]{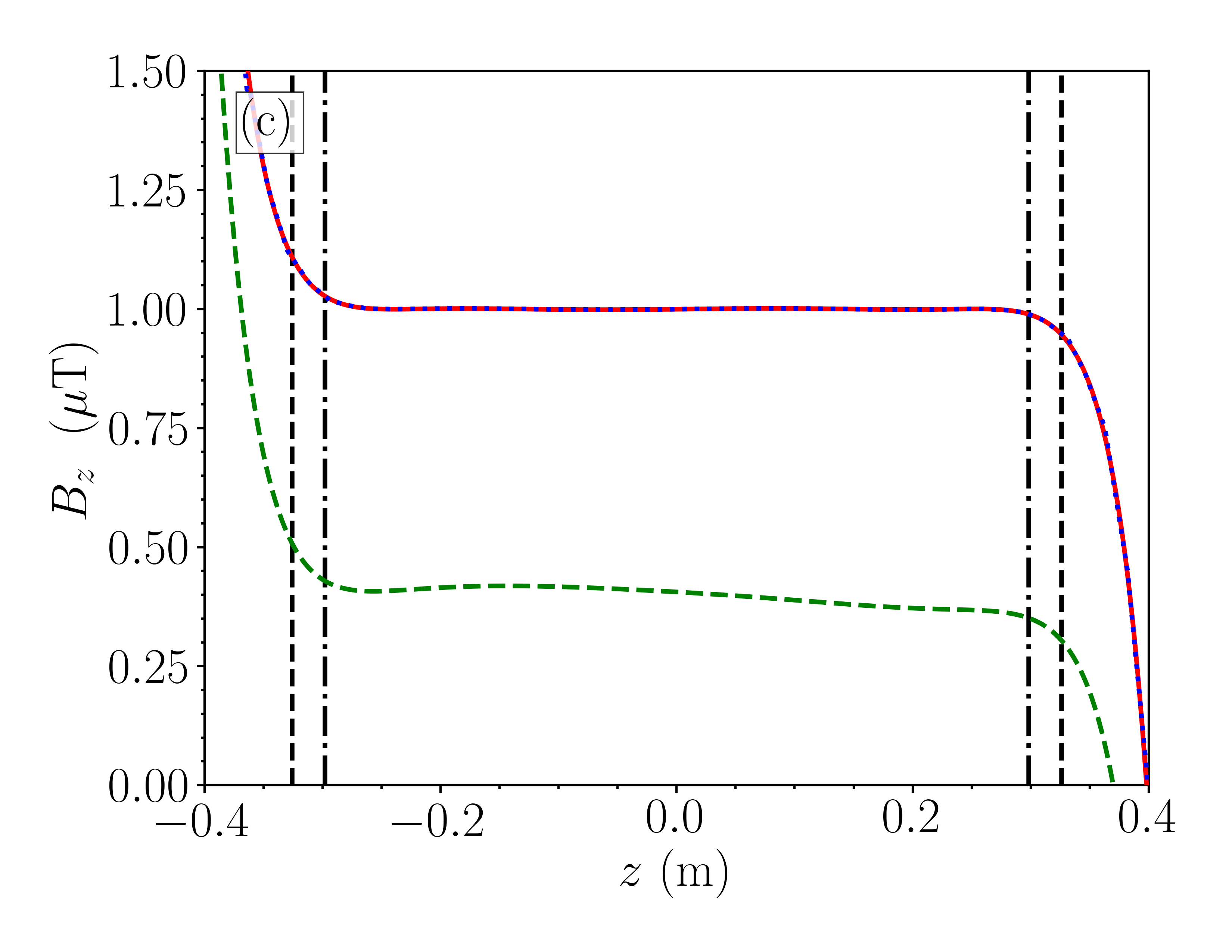}
         \caption{}
         \label{fig.bz_bvz}
     \end{subfigure}
\end{center}
\caption{Wire layouts (a--b) and performance (c) of a hybrid active--passive linear system optimized to generate a constant axial field, $B_z$. Current flows on the surface of two disks of radii $\rho_{c1}=0.95$ m (on the upper current-carrying plane in Fig.~\ref{fig.shieldandcoil}) and $\rho_{c2}=0.35$ m (on the lower current-carrying plane in Fig.~\ref{fig.shieldandcoil}), respectively, which are separated symmetrically from the origin and lie at $z'=\pm0.45$ m. The wire layouts are optimized to generate a constant axial field, $B_0=1$~$\mu$T, along the cylinder and parallel to its axis of symmetry. The current-carrying planes are placed symmetrically inside a perfect closed magnetic shield of radius $\rho_s=1$ m and length $L_s=1$ m and the magnetic field is optimized along the $z$-axis between $z=\pm{z'/2}$. The least squares optimization was performed with parameters $N=100$, $M=0$, $\beta=1.77\times10^{-8}$~T\textsuperscript{2}/W, $t=0.5$~mm, and $\rho=1.68\times10^{-8}$~$\Omega$m. (a--b) Color maps of the optimal current streamfunction on the upper and lower planar disks, respectively [blue and red shaded regions correspond to current counter-flows and their intensity shows the streamfunction magnitude from low (white) to high (intense color)]. Solid and dashed black curves represent discrete wires with opposite senses of current flow, approximating the current continuum with $N_\varphi=6$ global contour levels. (c) Axial magnetic field, $B_z$, versus axial position, $z$, calculated from the current continuum in (a--b) in three ways: analytically using \eqref{eq.cbrff}-\eqref{eq.cbzff} (solid red curve); numerically using COMSOL Multiphysics\textsuperscript{\textregistered} Version 5.5a, modeling the high-permeability cylinder as a perfect magnetic conductor (blue dotted curve); numerically \emph{without} the high-permeability cylinder and using the Biot–Savart law with $N_\varphi=100$ contour levels (dashed green curve). Black lines enclose the regions where the axial field gradient deviates by $5$\% (dashed) and $1$\% (dot-dashed).}
\label{fig.bzcoilnew}
\end{figure}
To further validate our design process, in Figs.~\ref{fig.dbxcoil}--\ref{fig.bzcoilnew} we show additional bi-planar hybrid active–passive systems and analyze their behavior. These systems are designed using our open-access Python code \cite{pjpython}. The coordinate axes and magnetic field plots are labeled in the same way as the systems presented in the main text. For both systems, the coils lie in the $z'=\pm 0.45$ m planes, as in the main text and shown in Fig.~\ref{fig.shieldandcoil}, and the fields are optimized along the $z$-axis between $z=\pm z'/2$. The design in Fig.~\ref{fig.dbxcoil} generates linear transverse field gradient, $\mathrm{d}B_x/\mathrm{d}z$, whose spatial variation matches the harmonic $\mathbf{B}=(z~\boldsymbol{\hat{x}} + x~\boldsymbol{\hat{z}})$ between two identical plates of radius $\rho_c=0.45$ m, inside a magnetic shield of radius $\rho_s=0.5$ m and length $L_s=1$ m. The field gradient is amplified by a factor of $1.06$ at the shield's center. If the shield is lengthened, the field gradient is diminished to $0.81$ at the shield's center in the long shield limit, as expected from our analysis in the main body of the text. The design in Fig.~\ref{fig.bzcoilnew} generates a uniform field, $B_z$, in the $z$-direction and extending between two planes of different sizes, with upper and lower coil sizes $\rho_{c1}=0.95$ m and $\rho_{c2}=0.35$ m, respectively, inside a magnetic shield of radius $\rho_s=1$ m and length $L_s=1$ m. Since the shield and coils have a relatively wide form factor, the field is highly uniform, with the region of axial field deviation below $1$\% extending over more than $70$\% of the distance between the planes.

\bibliographystyle{aipsamp}

\bibliographystyle{unsrt}


\end{document}